\documentclass[10pt,journal,compsoc]{IEEEtran}
\IEEEoverridecommandlockouts

\usepackage [nocompress]{cite}           

\usepackage[cmex10]{amsmath}
\usepackage{amssymb}
\usepackage{stackrel}
\usepackage{pifont}
\usepackage {amsthm}
\usepackage{graphicx}
\usepackage{tikz}
\usetikzlibrary{%
  calc,
  arrows,%
  shapes.misc,
  shapes.arrows,%
  chains,%
  matrix,%
  positioning,
  scopes,%
  decorations.pathmorphing,
  shadows,%
  decorations.text
}

\usepackage{pgfplots}
\pgfplotsset{compat=newest}

\usepackage{color,soul}
\usepackage{longtable}

\usepackage{subfigure}

\usepackage[]{algorithm2e}
\usepackage {enumitem}
\newtheorem{defi}{Definition}
\newtheorem{theorem}{Theorem}
\newtheorem{lemma}{Lemma}
\newtheorem{remark}{Remark}

\newtheorem{example}{Example}

\newcommand{\rom}[1]{\uppercase\expandafter{\romannumeral #1\relax}}
\newcommand{\Rom}[1]{\lowercase\expandafter{\romannumeral #1\relax}}
\renewcommand{\qed}{\hfill$\square$} 
\newcommand*\circled[1]{\tikz[baseline=(char.base)]{
            \node[shape=circle,draw,inner sep=2pt] (char) {#1};}}

\begin{document}

\title{Fundamental Limits of Private User Authentication}


\author
{
Narges~Kazempour, Mahtab~Mirmohseni and Mohammad~Reza~Aref 
\IEEEcompsocitemizethanks{\IEEEcompsocthanksitem N. Kazempour, M. Mirmohseni, and M. Aref are with the Information Systems and Security Lab (ISSL), Department of Electrical Engineering, Sharif University of Technology, Tehran, Iran,\protect\\
E-mail: Email: n.kazempour@ee.sharif.edu,\{mirmohseni,aref\}@sharif.edu
}

\thanks{The material in this paper has been  presented in part at ITW 2019 \cite{DBLP:conf/itw/KazempourMA19}.}}



\IEEEtitleabstractindextext{
\begin{abstract}
Most of the security services in the connected world of cyber-physical systems necessitate authenticating a large number of nodes privately. In this paper, the private authentication problem is considered which consists of a certificate authority, a verifier (or some verifiers), many legitimate users (provers), and an arbitrary number of attackers. Each legitimate user wants to be authenticated (using his personal key) by the verifier(s), while simultaneously staying completely anonymous (even to the verifier).
On the other hand, an attacker must fail to be authenticated.
We analyze this problem from an information-theoretical perspective  and propose a general interactive information-theoretic model for the problem.
As a metric to measure the reliability, we consider the normalized total key rate whose maximization has a trade-off with establishing privacy.
The problem is considered in two different scenarios: single-server scenario (only one verifier is considered, which all the provers are  connected to) and multi-server scenario ($N$ verifiers are assumed, where each verifier is connected to a subset of users).
   For both scenarios, two regimes are considered: finite size regime (i.e., the variables are elements of a finite field) and asymptotic regime (i.e., the variables are considered to have large enough length). We propose achievable schemes that satisfy the completeness, soundness, and privacy properties in both single-server and multi-server scenarios in all cases. In the finite size regime, the main idea is to generate the authentication keys according to a secret sharing scheme. We show that the proposed scheme in the special case of multi-server authentication in the finite size regime is optimal.
In the asymptotic regime, we use a random binning based scheme that relies on the joint typicality to generate the authentication keys.
Moreover, providing the converse proof, we show that our scheme achieves capacity in the asymptotic regime both in the single-server and multi-server scenarios.   
 \end{abstract}

}
\maketitle
\IEEEdisplaynontitleabstractindextext

\IEEEraisesectionheading{\section{Introduction}
\label{sec:introduction}}

\IEEEPARstart{W}{ith} the growth of cyber-physical systems,  authentication is an essential security feature in communication systems.
Based on the application, authentication protocols may authenticate the user and/or the message.
In fact, an authentication protocol seeks the answer to who sends the data is the user authorized to access the service, or whether the message received is unaltered.
The traditional cryptography-based methods for user authentication are mostly based on key validation,  e.g., public key or password based authentication; while the signed hash of the message is traditionally used for the message authentication.

The nature of user authentication requires revealing the user's identity, at least partly. However, nowadays, by increasing privacy concerns, users wish to stay anonymous when sending authentication requests. Anonymity ensures that a user may access a resource or service without disclosing his identity. In other words, anonymity of a user means that the user is not identifiable (not uniquely characterized) within a specific set of users \cite{pfitzmann2008anonymity}. 
This has many emerging applications such as vehicular networks \cite{alexiou2013vespa}, cloud computing, distributed servers \cite{roman2013features}, crypto-currencies and services on blockchain \cite{van2013cryptonote}.
In this paper, we study the inherent contradiction between authentication and privacy, referred to as the private authentication (PA) problem, where a legitimate user while passing the authentication, does not reveal his identity even to the server (by hiding in a group of users), but an illegitimate user (attacker) fails to be authenticated.
In practice, the PA problem is encountered when the server should not distinguish the one who requests the service with which key is authenticated~\cite{alexiou2013vespa}. There are cryptographic methods such as algorithms based on zero-knowledge proof \cite{DBLP:journals/siamcomp/GoldwasserMR89}, using tokens for private authentication \cite{DBLP:conf/icdcs/ChenC15}, and other cryptographic protocols, known as privacy-aware or anonymous authentication \cite{tsai2015privacy,he2017anonymous}. However, to the best of our knowledge, the fundamental limits of the PA problem have not been studied. 

With no privacy concern, the authentication problem has been studied fundamentally in works such as \cite{simmons1984authentication,lai2009authentication,maurer2000authentication,xiao2008using,tu2018keyless}. In \cite{simmons1984authentication}, Simmons considers message authentication over a noiseless channel with a shared key between peers of authentication,  where the lower bounds on the success probability of impersonation attack and substitution attack are established.
In \cite{lai2009authentication}, the results of \cite{simmons1984authentication}  have been extended to noisy channels.
In \cite{maurer2000authentication}, hypothesis testing is proposed to derive generalized lower bounds on the success probability of impersonation and substitution attacks.
The idea of using characteristics of the wireless channel for authentication, instead of shared key, has been analyzed in works such as  \cite{xiao2008using,tu2018keyless}.

Privacy constraints from an information theory perspective have been studied in some problems\cite{sun2017capacity,sankar2013utility,lai2008privacy,ignatenko2009biometric,DBLP:conf/isit/WillemsI12}. With no connection to the authentication problem, \cite{sun2017capacity} derives the capacity of private information retrieval, and \cite{sankar2013utility} studies privacy in database systems.
The privacy leakage rate is studied in works such as  \cite{lai2008privacy,ignatenko2009biometric,DBLP:conf/isit/WillemsI12} (biometric security systems).
In biometric security systems, a user wishes to authenticate himself by some biometric characteristics, such as fingerprint, called enrollment sequence.
The privacy constraint is to avoid leakage of information about the  enrollment sequence. \cite{lai2008privacy,ignatenko2009biometric,DBLP:conf/isit/WillemsI12}
derive the trade-off between  the secret key rate, the secret key is generated from the enrollment sequence, and the privacy leakage.

\textbf{Our contributions.} In this paper, we propose an information-theoretic interactive setup for the user-authentication problem with privacy constraint.
We consider a PA problem consisting of a certificate authority (CA), a single verifier or multiple verifiers (e.g., servers), many legitimate users, and an arbitrary number of attackers.
A legitimate user wishes to be authenticated to the verifier  to gain access to a service, while he wants to stay anonymous to anyone observing the authentication process (in particular the verifier). 
An attacker wants to impersonate himself as a legitimate user and  gain unauthorized access to the service. The verifier(s) wants to understand that the user, who requests authentication (prover), either is  legitimate or  an attacker. We assume semi-honest verifiers; they follow the protocol but they are curious about the identity of the provers.
CA is an entity that shares correlated randomness between the verifier(s) and the legitimate users. This randomness can be used by the verifier(s) in the process of authentication to distinguish an attacker from a legitimate user.
The case of multiple verifiers models the scenarios where a user wishes to access different services from different servers using only one single key \cite{DBLP:journals/mta/RehmanGCAN21}.

In an authentication protocol, two conditions should be considered: the legitimate user who follows the protocol must pass the authentication (completeness condition), and an attacker must fail to authenticate (soundness condition). 
In a private authentication protocol, the privacy condition must also be satisfied meaning that the legitimate user stays anonymous, i.e., the verifier(s) cannot distinguish the legitimate users' identities. In addition, it is desirable that the exposure of the key of a legitimate user releases as little information about the other keys as possible. Thus, we consider a reliability metric defined as the normalized total key rate of the users. The higher the normalized total key rate, the more reliable the private authentication protocol.

As a solution to the above problem, we propose general interactive information-theoretic protocols. 
Our proposed PA protocols are considered in two scenarios: single-server and multi-server scenarios. In the single-server scenario, only one verifier is assumed, connected to all users. In the multi-server scenario, $N$ verifiers are assumed, such that a subset of the users are connected to each verifier. 
For this scenario, we study two cases: ($\text{\Rom{1}}$) Individual authentication, where each verifier privately authenticates the users who are connected to him (same as the single-server scenario). ($\text{\Rom{2}}$) Distributed authentication, where the process of privately authenticating the user is done distributively by the verifiers that are connected to the user, by collaboratively sending the helper data. 

Both single-server and multi-server scenarios are analyzed in two regimes: finite size and asymptotic regimes. In the finite size regime, the variables used in the PA protocol are elements of a finite field.  So, the inverse of the length of the key vector of each user is considered as the metric, in addition to the normalized key rate.
In the asymptotic regime, the variables of the protocol are sequences of the length of order $l$, for arbitrary large $l$, where $l$ is the length of the key available at the legitimate user, required for authentication.

In the single-server scenario, for both the finite size and asymptotic regimes, we propose an achievable PA protocol, that satisfies the completeness, soundness, and privacy conditions.
In the asymptotic regime, the proposed scheme achieves capacity, proving its optimality.
The privacy condition is guaranteed by generating and extracting correlation between the information distributed among the users, which enables us to generate correlated keys such that checking the availability of the keys at the users does not need the identity of the user.
Our  finite size scheme mainly uses the idea that the keys of the legitimate users should lie on a specific polynomial. Checking that this polynomial passes through the key of the prover without knowing the exact value of the key, satisfies privacy constraint.
Here, the achievable region (normalized total key rate and inverse of the length of the key vector) is provided.
For the asymptotic regime, we propose an optimal scheme, where the main idea is to use the random binning and joint typicality between the keys of the legitimate users and the data available at the verifier to generate the required correlation. Using this correlation enables us to guarantee privacy.

In the multi-server scenario, for the individual authentication case, in both the finite size and asymptotic regimes, we propose achievable protocols. 
Both these protocols are the generalization of the protocols proposed in the single-server scenario, and thus have their properties. The proposed protocol in the asymptotic regime is optimal.
For the distributed authentication case, assuming that all the verifiers are connected to all the provers, we propose an optimal scheme in the finite size regime.
 This protocol uses symmetric private information retrieval (SPIR) \cite{DBLP:journals/tit/SunJ19} at the verifiers to send the desired data by the prover privately, and simultaneously the prover cannot gain information about other users' keys. In this protocol, the authentication process is done distributively by the verifiers.

\textbf{Organization.}  Section~\ref{sysmod} presents the system framework.  Section~\ref{sec::results} describes the results.
Section~\ref{sec::proofs} proposes the achievable schemes. 
Section~\ref{Sec::Analyze&comp} concludes the paper and provides a comparison of proposed authentication protocols.

\textit{Notations}: Capital letters are used to show random variables and small letters are their realizations.
The mutual information between $X$ and $Y$ is shown with $I(X;Y)$,  $H(X)$ shows the entropy of discrete random variable $X$.
 $\mathcal{T}_{\xi}^{(l)} (X,Y)$ is the set of strong jointly typical sequences $(X^l,Y^l)$.
 $[K]=\{1, 2, \cdots, K\}$, $X_{1:N}=\{X_1, \cdots, X_N\}$ and $X\sim Y$ means that $X$ and $Y$ are identically distributed.
 $P\left(n,k\right)=\frac{n!}{(n-k)!}$ shows $k$ permutations of $n$. For stating asymptotic results (Landau notation), $f(x)=\Theta(g(x))$ if $\lim_{x\to \infty} \frac{f(x)}{g(x)} <\infty$.

\section{Problem statement} \label{sysmod}
The PA framework consists of a CA, single or multiple verifiers, $K$ legitimate users, and an attacker\footnote{Arbitrary number of attackers may exists, but considering one of them is enough.}.
Each legitimate user wants to be authenticated without revealing his identity to the verifier(s).
Two scenarios are considered: Single-server scenario (one verifier) and multi-server scenario ($N$ verifiers).
 We first introduce the single-server scenario.
 \begin{figure}
\centering
\scalebox{0.6}{%
\begin{tikzpicture} [
        point/.style={coordinate},>=stealth',thick,draw=black,
        tip/.style={->,shorten >=0.007pt},
        text height=1.5ex,text depth=.25ex 
    ]

\draw[fill=cyan] (4.5,9) rectangle (7.5,10.5) node[pos=.5]{CA};
\draw[fill=teal!50] (0,6.5) rectangle (3,8) node[pos=.5]{Verifier};
\draw[fill=green!50] (9,6.5) rectangle (12,8) node[pos=.5]{Prover (user $k$)};

\path (6,10.5) node (CA) {};
\path (6,11.4) node (W) {$\mathcal{W}$};
\draw [->] (W) --(CA);

\path (1.5,8)	node (V) {};
\path (10.5,8)	node (p) {};
\path (4.5,9.75)	node (c1) {};
\path (7.5,9.75)	node (c2) {};
\draw [->] (c1) -- node [above, midway] {$V$} (V);
\path (2.5,9) node (aaaa) {\circled{3}};
\draw [->] (c2) -- node [above, midway] {$C_k$ } (p);
\path (9.5,9) node (aaa) {\circled{1}};

\draw [double,->] (9,5.5) -- node [above, midway] {$Q^{[k]}$} (3,5.5);
\path (6.8,6) node (aaaa) {\circled{2}};
\path (1.5,4.5)	node (v1) {$(S^{[k]},M^{[k]})=f^{[k]}(V,Q^{[k]})$};
\path (3.6,4.5)	node (v2) {};
\path (8.4,4.5)	node (p1) {};
\draw [double,->] (v2) -- node [above, midway] {$M^{[k]}$} (p1);
\path (6.8,5) node (aaaa) {\circled{4}};
\path (10.5,4)	node (p2) {$\hat{S}^{[k]}=g^{[k]}(C_k,M^{[k]},Q^{[k]})$};
\path (9,3.5)	node (p3) {};
\path (3,3.5)	node (v3) {};
\draw [double,->] (p3) -- node [above, midway] {$\hat{S}^{[k]}$} (v3);
\path (6.8,4) node (aaaa) {\circled{5}};
\path (1.5,3) node (v4) {accept if $\hat{S}^{[k]}=S^{[k]}$};
\path (1.5,2.5) node (v4) {reject if $\hat{S}^{[k]} \neq S^{[k]}$};

\end{tikzpicture}
%
%
%
%
%
%
%
%
%
}
\caption{Single-server PA protocol. In the key distribution phase, CA generates  $C\!\!=\!\!\{C_1, \cdots, C_K\}\! \in\! \mathcal{C}^K$; he sends  $C_k$ to the user $k$. In the authentication phase, after receiving the authentication query from the prover, CA generates $V$ and sends $V$ to the verifier.
The verifier generates $S^{[k]}$ as a secret and $M^{[k]}$ as a helper data. He sends $M^{[k]}$ to the prover. If the prover is  legitimate , he uses $(\!M^{[k]},C_k, Q^{[k]})$ to compute $\hat{S}^{[k]}$ and sends it to the verifier. If the received secret at the verifier is equal to $S^{[k]}$, the prover passes  authentication, unless he fails.}
\label{protocol}
\end{figure}  

\subsection{Single-server (verifier)}
A single-server PA protocol consists of a CA, a verifier, $K$ legitimate users, and an attacker. 
Each legitimate user wishes to authenticate himself to the verifier without revealing his identity.
The PA protocol, shown in Fig.~\ref{protocol}, has two phases: \\
1) \textit{Key distribution phase}: In this phase, the CA having access to $\mathcal{W}$, generates   $C_k\in \mathcal{C}$ as the personal key of user $k$ using function $h^{[k]}: \mathcal{W}\to \mathcal{C}$, where $\mathcal{W}$ is an authentication space, and sets 
$C=\left(C_1, \cdots, C_K\right) \in \mathcal{C}^K$. CA sends $C_k$ to the legitimate user $k$, which will be used later by the verifier to authenticate user $k$. This phase runs only once at the setup of the protocol to provide the users with their personal keys.\\
2) \textit{Authentication phase}: 
Here, the user $k \in [K]$ wants to prove knowledge of $C_k \in C$ without revealing index $k$ to the verifier. 
The decoding and encoding functions of the protocol must be designed to specify 
the verification process. The set of variables for a PA protocol is shown as $\mathcal{P}^{\text{Single}}=\left(V, Q^{[k]}, M^{[k]}, S^{[k]}, \hat{S}^{[k]} \right)$.  These parameters will be defined in the following.
Any time that a prover requests authentication from the verifier, this phase runs.
A prover is a user who sends the authentication request to the verifier. 
We have two cases:
Case $\text{\rom{1}}$: $\mathcal{H}=0$, the prover is a legitimate user,  Case $\text{\rom{2}}$: $\mathcal{H}=1$, the prover is an attacker.
When the prover claims that he has some $\bar{C} \in C$, the verifier decides on $\hat{\mathcal{H}}$, where   $\hat{\mathcal{H}}=0$ means that the verifier accepts the prover as a legitimate user, and $\hat{\mathcal{H}}=1$ means that the verifier rejects the prover.
The authentication phase, as shown in Fig.~\ref{protocol}, consists of five steps: \\
\textbf{Case} $\text{\rom{1}}$: $\mathcal{H}=0$

\textbf{S1:} The prover (user $k$) sends authentication query $Q^{[k]}\in \mathcal{Q}$ to the verifier, a request for authentication. 

\textbf{S2:} CA generates $V \in \mathcal{V}$ using encoding function $f^c: \mathcal{W}\to \mathcal{V}$ and sends $V$ to the verifier.

\textbf{S3:} The verifier, knowing $V$ and receiving $Q^{[k]}$, uses encoding function $f^{[k]}:\mathcal{V}\times \mathcal{Q} \to \mathcal{S}\times \mathcal{M}$ to generate $S^{[k]}$ and $M^{[k]}$, i.e., $(S^{[k]},M^{[k]})=f^{[k]}(V,Q^{[k]})$. The output of the encoding function has two parts. The first part, $S^{[k]}$, is used as a secret in the process of authentication in the following steps and the second part, $M^{[k]}$, is used as helper data for the prover. Then, the verifier keeps $S^{[k]}$ and sends $M^{[k]}$ to the prover.

\textbf{S4:}  The prover (legitimate user $k$) computes $\hat{S}^{[k]}=g^{[k]}(M^{[k]},C_k,Q^{[k]})$, where $g^{[k]}: \mathcal{M} \times \mathcal{C} \times \mathcal{Q} \to \mathcal{S}$ is a decoding function, and sends $\hat{S}^{[k]}$ to the verifier. 

\textbf{S5:}  If $\hat{S}^{[k]}=S^{[k]}$, the verifier accepts the prover and announces $\hat{\mathcal{H}}=0$. Otherwise, he rejects the prover and announces $\hat{\mathcal{H}}=1$. \label{step5}\\
\textbf{Case} $\text{\rom{2}}$: $\mathcal{H}=1$\\
Since neither the verifier nor the CA has knowledge of $\mathcal{H}$ at the beginning of the authentication phase, the steps of the authentication phase are similar to that of case $\text{\rom{1}}$, except in Step 4. In Step 4, the prover (an attacker) computes $\hat{S}=g(M,Q)$, since the only information available at the attacker is $M$ and $Q$. Function $g$ can be any arbitrary decoding function such that $g: \mathcal{M} \times \mathcal{Q} \to \mathcal{S}$. The prover sends $\hat{S}$ to the verifier.

The PA protocol should satisfy three constraints: completeness, soundness, and privacy.\\
The completeness property assures that a legitimate user is accepted. We define the $P_{e1}$-completeness constraint as if the prover is the legitimate user, then the verifier announces $\hat{\mathcal{H}}=0$ with probability more than $1-P_{e1}$. Thus,
\begin{align}
P_{e1}\text{-completeness:} \quad \mathbb{P}[\hat{\mathcal{H}}=1|\mathcal{H}=0] \leq P_{e1}. \label{eq:compeleteness}
\end{align}
The soundness property guarantees that an attacker is rejected. The $P_{e2}$-soundness constraint is defined as if the prover is an attacker, then the verifier announces $\hat{\mathcal{H}}=1$ with probability more than $1-P_{e2}$. Thus,
\begin{align}
P_{e2}\text{-soundness:} \quad \mathbb{P}[\hat{\mathcal{H}}=0|\mathcal{H}=1] \leq P_{e2}. \label{eq:soundness}
\end{align}
To make user identity private, all provers' requests  should be indistinguishable from the verifier's perspective, i.e., knowing the variables available at the verifier, all requests must be identically distributed. Thus, perfect privacy is defined as:
\begin{align*}
(\!V,Q^{[1]},S^{[1]},M^{[1]},\hat{S}^{[1]}\!)\!\sim\! (\!V,Q^{[k]},S^{[k]},M^{[k]},\hat{S}^{[k]}\!), \, \forall k\! \in\! [K]. 
\end{align*}
In the proposed model, we use the $P_p$-privacy property as:
\begin{align}
P_p
&\text{-privacy:} \quad \mathbb{P}\left[ (V,Q^{[1]},S^{[1]},M^{[1]},\hat{S}^{[1]})\sim\right. \nonumber \\
&\left.(V,Q^{[k]},S^{[k]},M^{[k]},\hat{S}^{[k]}), \quad \forall k \in [K] \right] \geq 1-P_p. \label{eq:privacy}
\end{align}
Considering $\text{length}(C)$ as the length of the keys, we define a reliability metric, $R$, as the normalized total key rate:
\begin{align}
R=\frac{H(C_1,C_2,\cdots, C_K)}{K \text{length}(C)}. \label{eq:rate}
\end{align}
Since $C_k$ is the personal key of user $k$, it is desirable that $(C_1,C_2,\cdots, C_K)$ has the maximum entropy, so that exposure of a user's key reveals minimum information about other users' keys.
In other words, users wish that others gain as little information as possible about their keys.
 On the other hand, for preserving privacy, it is desired that the personal keys of the users have correlated information. This correlated information is used for authentication without revealing the identity of the user. Thus, there is a trade-off between maximizing the entropy of $(C_1,C_2,\cdots, C_K)$ and establishing privacy.\\
The goal is to design a PA protocol that satisfies completeness, soundness, and privacy constraints while achieving maximum $R$.
Two regimes are considered: finite size and asymptotic regimes\footnote{Both finite size and asymptotic regimes are considered in terms of size of the variable and both of them are single round.}, defined in the following.

\textbf{Finite size regime:}
In the finite size regime, the variables of the protocol are chosen as elements of a finite field. More precisely, $\mathcal{W}=GF(q^L)$, $\mathcal{S}= GF(q^L)$, $\mathcal{C}=(GF(q^L))^{l_c}$, $\mathcal{V}= (GF(q^L))^{l_v}$ and $\mathcal{M}= (GF(q^L))^{l_m} $, where $q$ is a prime number and $L$, $l_c$, $l_v$, and $l_m$ are integers\footnote{$\mathcal{S}= GF(q^L)^{l_s}$ is feasible, but to avoid complexity, we set $l_s=1$.}.
For example, for a variable $X$ used in the protocol, $X \in \left(GF(q^L)\right)^{l_x}$, meaning that $X$ is a $l_x$-tuple where each element is of length $L$ in a $q$-ary unit. $l_x$ will be chosen properly according to the protocol.
 Thus, $\text{length}(X)=l_x L$.
All operations in this regime are done in the finite field $GF(q^L)$.
Thus, $C_k\in (GF(q^L))^{l_c}$ for $k\in[K]$. We also define the inverse of the length of the key vector as:
\begin{align*}
R_c=\frac{1}{l_c}.
\end{align*}
\begin{defi} \label{def:R_fin}
The pair $(R,R_c)$ is achievable for a PA protocol in the finite size regime, if there exist encoding functions $f^c$ and $f^{[k]}$, decoding functions $g^{[k]}$ and  functions $h^{[k]}$ for $k \in [K]$, such that $P_{e1}= 0$, $P_{e2} = \frac{1}{|\mathcal{S}|}$ and $ P_p =0$. It must be noted that since $l_c\in \mathbb{Z}^+$, the achievable region is countable. 
\end{defi}
\begin{remark}
It is worth noting that, since an attacker can guess the key, the lowest possible $P_{e2}$ is equal to $\frac{1}{|\mathcal{S}|}$.
\end{remark}
\begin{defi} \label{def:C_fin}
The capacity region $\mathfrak{C}^{\text{fin}}$ for a PA protocol is  the union  of all achievable pairs.
\end{defi}
\textbf{Asymptotic regime:}
In the asymptotic regime, the elements of the protocol are sequences of arbitrary large length. More exactly, each variable of the protocol, e.g., $X^{l_x}$ is a sequence of length $l_x$ with each component chosen from $\mathcal{X}$. Thus, $\text{length}(X^{l_x})=l_x$, where $l_x$ approaches infinity. In particular, considering $\mathcal{C}=\mathcal{Y}^l$, we define the $l$-length key $C_k=Y_k^l=\left({Y_k}_1,\cdots, {Y_k}_l\right)$ where ${Y_k}_i$ is a random variable from alphabet $\mathcal{Y}$, $k\in [K]$ and $i\in \{1, 2,\cdots, l\}$.
\begin{defi} \label{def:R_asymp}
The normalized total key rate $R$ is achievable for a PA protocol in the asymptotic regime, if there exist encoding functions $f^c$ and $f^{[k]}$, decoding functions $g^{[k]}$ and  functions $h^{[k]}$ for $k \in [K]$, such that $P_{e1}$, $P_{e2} $ and $ P_p $ tend to zero as the length of the variables goes to infinity.
\end{defi}

\begin{defi} \label{def:C_asymp}
The key capacity for a PA protocol in the asymptotic regime is:
\begin{align*}
\mathfrak{C}^{\text{asy}} \triangleq \sup \{R: R &\text{ is an achievable normalized total key rate} \nonumber \\
&\text{ for a PA protocol in asymptotic regime}\}.
\end{align*}
\end{defi}
\subsection{Multi-server (verifiers)}
In this section, a multi-server version of the PA framework is introduced.
The multi-server PA protocol, shown in Fig.~\ref{fig::distpa}, consists of a CA, $N$ verifiers, $K$ legitimate users, and an attacker.
Each legitimate user wants to gain access to a subset of verifiers using only one personal key.  Verifier $n$ ($n \in [N]$) must authenticate any legitimate user in the set  $\mathcal{N}_n \subseteq [K]$. $\mathcal{N}_n$ can be any arbitrary subset of $[K]$ and shows the users that are allowed to connect to the verifier $n$. However, for privacy reasons, the user wants to be anonymous in the set $\mathcal{N}_n$, that is, the verifier $n$ should not understand which user in the set  $\mathcal{N}_n$ requests accessing the service.
\begin{figure}
\centering
\scalebox{0.5}[0.4]{%
\begin{tikzpicture} [
        point/.style={coordinate},>=stealth',thick,draw=black,
        tip/.style={->,shorten >=0.007pt},
        text height=1.5ex,text depth=.25ex, every node/.style={scale=1.5} 
    ]

\draw[fill=cyan] (4.5,9) rectangle (7.5,10) node[pos=.5]{CA};
\draw[fill=teal!50] (9,7) rectangle (12,8) node[pos=.5]{Verifier $1$};
\draw[fill=teal!50] (9,3.25) rectangle (12,4.25) node[pos=.5]{Verifier $n$};
\path (10.5,5.625)	node (V) {$\vdots$};
\path (10.5,1.875)	node (V) {$\vdots$};
\draw[fill=teal!50] (9,-0.5) rectangle (12,0.5) node[pos=.5]{Verifier $N$};
\draw [decorate,decoration={brace,amplitude=4pt},xshift=0.5cm,yshift=0pt]     (3,6.5) -- (3,0.5) node [midway,right,xshift=.1cm,yshift=0pt] {$\mathcal{N}_n$}; 
\draw [purple,decorate,decoration={brace,amplitude=4pt,mirror},xshift=-0.5cm,yshift=0pt]     (9,6) -- (9,-0.5) node [midway,left,xshift=-.1cm,yshift=-16pt] {$\mathcal{N}^{[k]}$}; 

\draw[fill=green!50] (0,7) rectangle (3,8) node[pos=.5]{User $1$};
\draw[fill=green!50] (0,5.5) rectangle (3,6.5) node[pos=.5]{User $2$};
\path (1.5,4.5)	node (V1) {$\vdots$};
\path (1.5,1.5)	node (V1) {$\vdots$};
\draw[fill=green!50] (0,2.5) rectangle (3,3.5) node[pos=.5]{User $k$};
\draw[fill=green!50] (0,-0.5) rectangle (3,0.5) node[pos=.5]{User $K$};

\draw [dotted] (8.7,4.25) --  (3.8,6.5);
\draw [dotted] (8.7,3.25) --  (3.8,0.5);

\draw [purple,dashed] (3.3,3.5) --  (8.2,6);
\draw [purple,dashed] (3.3,2.5) --  (8.2,-0.5);

\node (A) at (4.5, 9.2) {};
\node (Av) at (4,9.5) {\small $\vdots$};
\node (AA) at (4.5, 9.8) {};
\node (B) at (1.5, 8) {};
\node (C) at (0, 0) {};
\node (D) at (10.5, 8) {};
\node (E) at (7.5, 9.2) {};
\node (Ev) at (8, 9.5) {\small $\vdots$}; 
\node (EE) at (7.5, 9.8) {}; 
\node (F) at (12, 0) {}; 
    
\draw[->, to path={-| (\tikztotarget)}]
  (A) edge (B)  (E) edge (D);  
  
  \path (A) -| node[pos=0.75,right] {$C_1$} (B); 
  \path (E) -| node[pos=0.75,left] {$V_1$} (D);  
  
  \draw[-stealth] (AA)
               -| (-1.5,9.5) |- (C); 
  \draw[-stealth] (EE)
               -| (13.5,9.5) |- (F);  
               
 \path (AA) -| node[pos=1,left] {$C_K$} (C); 
  \path (EE) -| node[pos=1,right] {$V_N$} (F); 
    
\end{tikzpicture}
}
\caption{Multi-server (verifiers) PA Protocol.}
\label{fig::distpa}
\end{figure} 
\begin{defi}
$(k,n)$ forms a feasible pair if $k\in \mathcal{N}_n$. This means that the user $k$ has the right to gain access to the verifier $n$.
\end{defi}
The same as the single-server scenario, the multi-server PA protocol has two phases:\\
1) \textit{Key distribution phase}: Similar to the single-server case, CA sends the personal key of user $k$, $C_k$, to each legitimate user.\\
2) \textit{Authentication phase}: The user $k$ wants to prove knowledge of $C_k \in C$ without revealing index $k$  to the verifier~$n^\ast$. Any time that a prover wants to be authenticated, this phase runs.
In this phase, the verification process is specified, which includes the variables $\mathcal{P}^{\text{Multi}}=\left( V_n, Q_n^{[k]}, M_n^{[k]}, S_n^{[k]}, \hat{S}_{n}^{[k]}, \, n \in [N] \right)$. These parameters will be defined in the following.
It must be noted that the flow of the PA protocol for the single-server and the multi-server is similar, this means that the one who takes action in each step of the authentication phase is the same. However, considering the type of authentication in multi-server scenarios, two situations are studied: 1) Individual authentication, 2) distributed authentication.
In these two situations, when helper data is sent, interactions between the parties of the protocol are different. The details are presented in the following in Step~3.
Because of the similarities  between the overall functionality of Case $\text{\rom{1}}$ (the prover is a legitimate user) and Case $\text{\rom{2}}$ (the prover is an attacker), only Case $\text{\rom{1}}$ is described. Denoting $\mathcal{N}^{[k]}$ as the set of verifiers that user $k$ forms a feasible pair with, the authentication phase is in Case $\text{\rom{1}}$ is as follows.

\textbf{S1:} The legitimate user $k$ sends authentication query $Q_n^{[k]}\in \mathcal{Q}$ to the verifier $n$, where $n\in \mathcal{N}^{[k]}$. This query consists of the request for authentication.

\textbf{S2:} CA generates $V_n \in \mathcal{V}$, $n\in[N]$ using the encoding function $f_n^c: \mathcal{W}\to \mathcal{V}$, where $\mathcal{W}$ is an authentication space. CA sets $V=\{V_1, V_2, \cdots, V_N\}$ and sends $V_n$ to the $n$-th verifier.

\textbf{S3:} Two situations are considered:
\begin{itemize}
\item Individual authentication: only verifier $n^\ast$ sends helper data. In the individual authentication, the $n^\ast$-th verifier, knowing $V_{n^\ast}$, uses the encoding function $f_{n^\ast}^{[k]}$ to generate $S_{n^\ast}^{[k]}$ and $M_{n^\ast}^{[k]}$, i.e., $(S_{n^\ast}^{[k]},M_{n^\ast}^{[k]})=f_{n^\ast}^{[k]}(V_{n^\ast},Q_{n^\ast}^{[k]})$ and  $f_{n^\ast}^{[k]}:\mathcal{V} \times \mathcal{Q} \to \mathcal{S}\times \mathcal{M}$. The verifier keeps $S_{n^\ast}^{[k]}$ and sends $M_{n^\ast}^{[k]}$ to the prover.
\item Distributed authentication\footnote{It must be mentioned that distributed authentication is the general case of multi-server authentication and individual authentication is a special case of distributed authentication where only verifier $n^\ast$ sends helper data.}: all the verifiers in $\mathcal{N}^{[k]}$ can send helper data to the user $k$. In the distributed authentication, for each $n\in \mathcal{N}^{[k]}\backslash \{n^\ast\}$, the $n$-th verifier uses the encoding function $f_n^{[k]}$ to generate $M_n^{[k]}$, i.e., $M_{n}^{[k]}=f_n^{[k]}(V_n,Q_n^{[k]})$ and  $f_n^{[k]}:\mathcal{V} \times \mathcal{Q} \to \mathcal{M}$. 
Also, the verifier $n^\ast$ generates $S_{n^\ast}^{[k]}$ and $M_{n^\ast}^{[k]}$, i.e., $(S_{n^\ast}^{[k]},M_{n^\ast}^{[k]})=f_{n^\ast}^{[k]}(V_{n^\ast},Q_{n^\ast}^{[k]})$. For $n\in \mathcal{N}^{[k]}$, each verifier $n$ sends $M_n^{[k]}$ to the prover, and verifier $n^\ast$ keeps $S_{n^\ast}^{[k]}$ to be used in the following steps for checking the legality of the prover.
\end{itemize}

\textbf{S4:}  The prover (legitimate user $k$) computes $\hat{S}_{n^\ast}^{[k]}=g_{n^\ast}^{[k]}(C_k,M_{1:N}^{[k]},Q_{1:N}^{[k]})$, where $g_{n^\ast}^{[k]}$ is a decoding function and $g_{n^\ast}^{[k]}: \mathcal{M} \times \mathcal{C} \times \mathcal{Q}\to \mathcal{S}$. It must be emphasized that the prover can use the queries he sent to the verifiers he forms feasible pair with and the helper data he received from these verifiers to estimate the secret generated by a specific verifier.
 The prover sends $\hat{S}_{n^\ast}^{[k]}$ to the verifier $n^\ast$.

\textbf{S5:} If $\hat{S}_{n^\ast}^{[k]}=S_{n^\ast}^{[k]}$, the verifier accepts the prover and announces $\hat{\mathcal{H}}=0$; Otherwise, he rejects the prover and announces $\hat{\mathcal{H}}=1$. 

For each feasible pair,  $P_{e1}$-completeness constraint is defined as \eqref{eq:compeleteness}, and  for each verifier  $P_{e2}$-soundness constraint is defined as   \eqref{eq:soundness}. For verifier $n$, the privacy constraint is:
\begin{align}
&P_p\text{-privacy:} \quad \mathbb{P}\left[ (V_n,Q_n^{[k_1]},S_n^{[k_1]},M_n^{[k_1]},\hat{S}_{n}^{[k_1]})\sim \right. \nonumber \\
& \left.(V_n,Q_n^{[k]},S_n^{[k]},M_n^{[k]},\hat{S}_{n}^{[k]}), \quad \forall k,k_1 \in \mathcal{N}_n \right] \geq 1-P_p.  \label{eq::privacy_MS}
\end{align}
Also, the normalized total key rate is defined as \eqref{eq:rate}.\\
By designing the verification process for multi-server scenario, $\mathcal{P}^{\text{Multi}}$, the definitions of the achievable  pair and the capacity region in the finite size regime and the definitions of the achievable rate and the capacity in the asymptotic regime are the same as Definitions \ref{def:R_fin}-\ref{def:C_asymp}, respectively, considering appropriate encoding and decoding functions. 

\begin{remark}
In Subsection~\ref{subsec::multiserver_fin_thms}, these two different situations of individual authentication and distributed authentication are considered.
\end{remark}
%
%

\section{Results} \label{sec::results}
\subsection{Single-server, finite size regime} 
\begin{theorem} \label{finitethm}
The achievable region $(R,R_C)$ ($R$ in $q$-ary unit) for a single-server PA protocol in the finite size regime is 
\begin{equation}
\left\{\!\!
 \begin{array}{ll}
(R,R_c):\\
  R_c= 1; & R\leq \frac{1}{K}\\
 R_c=\frac{1}{i}, i \geq 2; & R\leq 1-R_c+\frac{R_c \log_q P(q^L-1,K)}{KL} 
\end{array}
\right\}. \label{eq:::thm_fin_single-pair}
\end{equation}
\end{theorem}

\begin{remark}
The achievable scheme is provided in Subsection~\ref{Subsec::SSFR}.
The achievability proof consists of two schemes, scheme 1 for $R_c=1$ and scheme 2 for $R_c= \frac{1}{i}, i\geq 2$. The idea for the first scheme is to share a common key between all legitimate users. The design of scheme 2 is depending on two facts: 1) a $k$ degree polynomial can cover $k+1$ randomly chosen points from  $GF(q^L)$,  2) $k+1$ points can uniquely determine a $k$ degree polynomial
 In this scheme, the CA generates $K$ random points (as the keys of the legitimate users) and computes the unique polynomial of degree $K$ passing through these points. The prover wishes to prove that his key is a valid point
on the polynomial $f(.)$ to confirm his legitimacy (inspired by the idea of Shamir's secret sharing (SSS) \cite{shamir1979share}). It must be emphasized that even though we use polynomial interpolation similar to SSS; however, we do not need to combine the shares (keys of the provers), and the setup is different.
\end{remark}
\begin{remark}
The normalized total key rate in \eqref{eq:::thm_fin_single-pair} tends to its upper bound ,$1$, as $q^L \to \infty$ for $l_c\geq 2$.
\end{remark}

\subsection{Single-server, asymptotic regime}
In the asymptotic regime, all the variables of the protocol are sequences of length $\Theta(l)$ for arbitrary large $l$.

\begin{theorem}  \label{asymptotic_thm}
The capacity of the single-server PA protocol in the asymptotic regime is equal to $\max_{p(y)} H(Y)$, where $Y$ is an arbitrary random variable chosen from $\mathcal{Y}$.
\end{theorem}
\begin{remark}
The achievable scheme is provided in Subsection~\ref{subsec::asympsingle_proof}.
The idea of the proposed scheme is to provide the verifier and the legitimate provers with correlated data (done by CA) using Wyner-Ziv coding \cite{DBLP:journals/tit/WynerZ76}, and then  verify the existence of this correlation. This verification is based on a random binning technique and is done in an interactive protocol. 
It must be noted that applying Wyner-Ziv coding should fit the framework of the problem, and we cannot use Wyner-Ziv coding, lonely.
\end{remark}

\begin{remark}
The proposed achievable scheme in Subsection~\ref{subsec::asympsingle_proof} is optimal.
 This means, we show that the normalized total key rate of the protocol achieves capacity.
\end{remark}
\subsection{Multi-server, finite size regime} \label{subsec::multiserver_fin_thms}
The results  provided in this subsection consider two different viewpoints of the problem of multi-server private authentication.
First, the individual authentication, where similar to the single-server case, each server tries to authenticate a group of users that it forms feasible pair with, by considering privacy constraint. In other words, in each verification process,  the legitimate user is connected to only one verifier, and that verifier privately authenticates the user with the corresponding key.
The second viewpoint is to use the multi-server structure to perform the verification process in a distributed manner. More precisely, unlike the individual authentication, all (or a subset) of the servers contribute in the process of privately authenticating the users.
\begin{theorem} \label{thm::Multi_server_fin}
The achievable region $(R,R_C)$ ($R$ in $q$-ary unit) for a multi-server PA protocol in the finite size regime is equal to \eqref{eq:::thm_fin_single-pair}.
\end{theorem}
\begin{remark}
The achievable scheme is proposed in Subsection~\ref{subsec::multi-indivi}, and is for the individual authentication case.
The proposed protocol consists of two schemes that are generalizations of the schemes proposed in the single-server case, Theorem~\ref{finitethm}.
Each verifier tries to privately authenticate the groups of users that it forms feasible pair with.
In scheme 1, a common key is shared between all users and verifiers. Each prover proves his legitimacy by providing this common key to the appropriate verifier.
In scheme 2, each verifier computes the unique polynomial that passes through the keys of the users that he forms a feasible pair with. By proving that his key lies on the polynomial constructed by the desired verifier, the prover convinces the verifier of his legitimacy.
\end{remark}

\begin{theorem} \label{thm::DPA}
When $\mathcal{N}_1=\mathcal{N}_2=\cdots=\mathcal{N}_n=[K]$, the capacity region of the multi-server PA protocol in finite size regime in $q$-ary unit is:
\begin{equation}
\left\{
 \begin{array}{ll}
(R,R_c):\\
R_c=\frac{1}{i}, i \geq 1; & R\leq 1
\end{array}
\right\}. \label{eq::thm_DPA}
\end{equation}
\end{theorem}

\begin{remark}
The achievable scheme is provided in Subsection~\ref{subsec::DPAproof}, and is for the distributed authentication case.
In the proposed protocol, the prover  asks for required data that fits his index,  from all verifiers. The queries are sent according to the SPIR protocol, and the verifiers sends answers  according to the  SPIR protocol. Using the SPIR protocol enables the prover to protect his privacy. The prover uses his own personal key and the retrieved data to estimate the secret. 
\end{remark}
\begin{remark}
The proposed achievable scheme for the distributed authentication is optimal and achieved capacity region. However, unlike the individual authentication, this scheme is provided for a special case where $\mathcal{N}_1=\mathcal{N}_2=\cdots=\mathcal{N}_n=[K]$.
\end{remark}
\begin{remark}
The proposed scheme for the distributed authentication provides some extra features compared with the individual authentication, that are discussed in Subsection~\ref{intro_extra}.
\end{remark}
\subsection{Multi-server, asymptotic regime}
Considering $C_k=Y_k^l$, we have the following theorem.
\begin{theorem} \label{thm::Multi_server_asymp}
The capacity of the multi-server PA protocol in the asymptotic regime  is equal to $\max_{p(y)} H(Y)$, where $Y$ is an arbitrary random variable chosen from $\mathcal{Y}$.
\end{theorem}
\begin{remark}
The proposed achievable scheme is provided in Subsection~\ref{subsec::multi-asys}, and is the generalization of the scheme proposed in Subsection \ref{subsec::asympsingle_proof}.
However, in the multi-server scenario each personal key must have the correlation properties with information available at all the verifiers he forms feasible pair with.
\end{remark}

\section{Proofs} \label{sec::proofs}
\subsection{Proof of Theorem \ref{finitethm} (single server, finite size regime)} \label{Subsec::SSFR}
\begin{figure}
\centering
\scalebox{0.8}{
\includegraphics[scale=0.65]{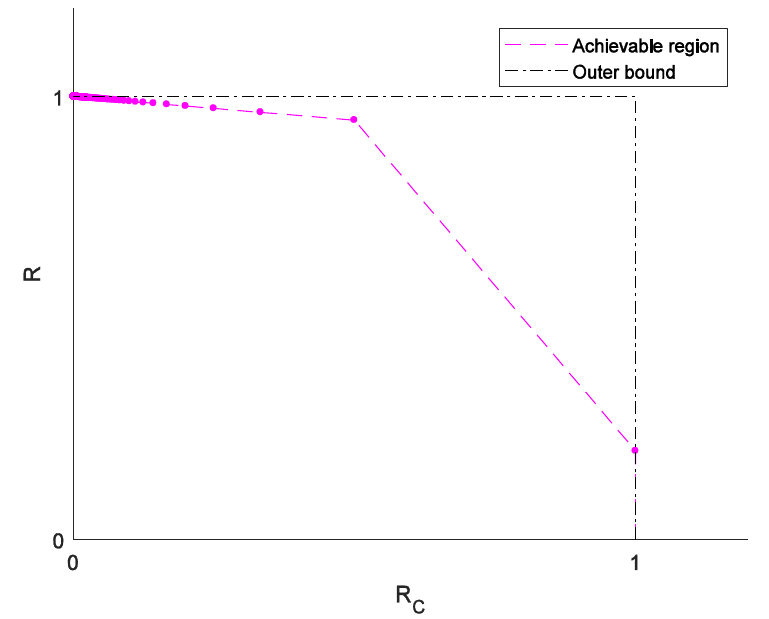}
}
\caption{(a) Achievable region and outer bound, Theorem~\ref{finitethm} and \ref{thm::Multi_server_fin}}
\label{fig::thm1}
\end{figure}

The achievability proof consists of two schemes, scheme~1 for $R_c=1$ and scheme 2 for $R_c= \frac{1}{i}, i\!\!\geq\!\! 2$. The achievable region, shown in Fig.~\ref{fig::thm1}, for the proposed protocol is the union of the achievable pairs results from these two schemes. 

\textbf{Scheme 1} for single-server, finite size regime (SS-FR1) protocol ($R_c=1, R\leq \frac{1}{K}$): \\
1) \textit{Key distribution phase}: first, CA chooses $\tilde{C} \in GF(q^L)$ uniformly at random and  sets $C_k=\tilde{C}$, $\forall k\in[K]$.\\
2) \textit{Authentication phase}: Considering user $k$ as the prover, the encoding and decoding functions follow:

\textbf{S1:} The prover (user $k$) sends the authentication query $Q^{[k]}$ to the verifier. It is worth noting that this query only consists of a request for authentication and does not ask for specific helper data.

\textbf{S2:} CA sets $V\!=\!\{\tilde{C}\}$, and sends it to the verifier. It is to be noted  since $V$ does not change in any verification process, $V$ can be sent to the verifier at the key distribution phase.

\textbf{S3:} The verifier sets $S^{[k]}=\tilde{C}$. It is obvious that since all the users have the secret as their own personal key, there is no need to helper data. Thus $M^{[k]}$ is Null.

\textbf{S4:} Having access to $C_k=\tilde{C}$, the prover sets $\hat{S}^{[k]}=\tilde{C}$ and sends it to the verifier.

\textbf{S5:} The verifier checks if $\hat{S}^{[k]}=S^{[k]}$, to pass or reject the prover.

These steps determine the variables of the verification process $\mathcal{P}^{\text{Single}}$.
Now we analyze the completeness, soundness,
 and privacy properties of this scheme.\\
 $\mathit{0}$\textit{-completeness}: Each legitimate user has access to $\tilde{C}$. Thus, user $k$ can estimate $\hat{S}^{[k]}=S^{[k]}=\tilde{C}$ for all $k\in [K]$ and thus:
\begin{align}
\mathbb{P}[\hat{\mathcal{H}}=1|\mathcal{H}=0]=\mathbb{P}[\hat{S}^{[k]}\neq S^{[k]}] =0. \label{eq::0complete_coomon_fr}
\end{align}
This proves the $0$-completeness property of the scheme.\\
$\mathit{\frac{1}{|\mathcal{S}|}}$\textit{-soundness}: Since  the helper data is null in this scheme, we have, $I(S;M^{[k]})=0$. This means that if the prover is an attacker, he obtains no information about the secret. This means that, the best possible attack strategy is to guess the secret.
Considering that $S$ is uniformly distributed):
\begin{align}
\mathbb{P}[\hat{\mathcal{H}}=0|\mathcal{H}=1]=\mathbb{P}[\hat{S}= S]=\frac{1}{q^L}. \label{eq::sound_common_fr}
\end{align} 
Thus, the maximum success probability of the attacker is equal to $\frac{1}{q^L}$, showing that the larger the size of the finite field, the lower the probability of attack. This completes the proof of $\frac{1}{|\mathcal{S}|}$-soundness property.\\
$\mathit{0}$\textit{-privacy}: As stated above, every legitimate user has access to $\tilde{C}$. Thus, $\hat{S}^{[1]}=\hat{S}^{[k]}=\tilde{C}$, and $(V,Q^{[1]},S^{[1]},M^{[1]},\hat{S}^{[1]})\sim(V,Q^{[k]},S^{[k]},M^{[k]},\hat{S}^{[k]})$ for all $k \in [K]$, and $0$-privacy is satisfied. Intuitively, since every legitimate prover always sends $\tilde{C}$ and authentication query is only a request for authentication (no difference between the legitimate users), the verifier cannot distinguish the identity of the prover.\\
Now, we compute the normalized total key rate of the proposed protocol as,
\begin{align}
R&=\frac{H(C_1,C_2,\cdots, C_K)}{K\text{length}(C)} =\frac{H(\tilde{C})}{Kl_cL} \stackrel{(a)}=\frac{H(\tilde{C})}{KL} =\frac{1}{K}, \label{R-common-fin}
\end{align}
where (a) follows due to $l_c=1$.\\
This scheme achieves the pair $(\frac{1}{K},1)$, shown  in Fig.~\ref{fig::thm1}.
We remark that, trivially, when $R=\frac{1}{K}$ is achievable, all rates less than $\frac{1}{K}$ can be achieved,
by changing the distribution of $\tilde{C}$ and thus reducing its randomness. This concept will be used in other schemes in the finite size regime, too.\\
\textbf{Scheme 2} for single-server, finite size regime (SS-FR2) protocol ($R_c=\frac{1}{i} (i\geq  2), R\leq 1-R_c+\frac{R_c}{KL} \log_q P(q^L-1,K)$): \\
1) \textit{Key distribution phase}: CA chooses $X_1, \cdots, X_K$ ($K$ distinct random numbers) from the finite field $GF(q^L)\backslash\{0\}$, and chooses $K$ variables from the field $\mathcal{W}=GF(q^L)$, independently and uniformly at random and sets them as $Y_1, \cdots, Y_K$.  Moreover, CA chooses independently and uniformly at random $K$  variables from $\left(GF(q^L)\right)^{l_c-2}$ and sets them as $Z_1,  \cdots, Z_K$.   CA sets $C_k=(X_k,Y_k,Z_k)$ and send $C_k$ to user $k$, $\forall k\in [K]$.\\
2) \textit{Authentication phase}: Each user wishes to prove that his key ($(X_k,Y_k)$) fits the polynomial.

\textbf{S1:} The prover sends authentication query $Q^{[k]}$ to the verifier. It is worth noting that this query only consists of a request for authentication.

\textbf{S2:} CA chooses $a_0$ from $GF(q^L)$ independently and uniformly at random, sets $V=\left\{a_0, C_k |\, k\in[K]\right\}$ and sends $V$ to the verifier.

\textbf{S3:} Having access to $V$, the verifier sets $S^{[k]}=a_0$. Since $K+1$ points can uniquely determine a $K$ degree polynomial, the verifier computes the unique polynomial $f(.)$ of degree $K$ that passes through $\left\{(0,a_0), (X_1,Y_1), \cdots , (X_K,Y_K)\right\}$. Then, the verifier chooses $K$ new points on the polynomial $f(.)$, i.e., the verifier chooses at random $K$ distinct elements $M_1, \cdots , M_K$ of $GF(q^L)\backslash\{0, X_k| \, k\in [K]\}$. The verifier sets $M^{[k]}=\left\{(M_i,f(M_i)), \, \forall i \in [K] \right\}$ and sends $M^{[k]}$ to user~$k$.

\textbf{S4:} Knowing $M^{[k]}$ and $C_k$, the prover has access to $K+1$ distinct points of the polynomial $f(.)$ (of degree~$K$),  thus, he can derive the polynomial $f(.)$ and compute $\hat{S}^{[k]}\!\!=\!\!f(0)\!\!=\!\!a_0$. The prover sends $\hat{S}^{[k]}$ to the verifier.

\textbf{S5:} The verifier checks if $\hat{S}^{[k]}=S^{[k]}$, to pass or reject the prover.\\
The above steps determine the variables of  $\mathcal{P}^{\text{Single}}$ of SS-FR2.
\begin{remark} \label{remark::constant a0}
It is to be emphasized that, $a_0$, selected in Step~2 of the authentication phase, must not be changed in different verification processes. Changing $a_0$ results in changing the polynomial $f(.)$, and the intersection points of the old and the new polynomials consist of the keys of the users. This violates security and reveals the keys of the users to other legitimate users.
Also, any time the verifier receives an authentication request, he must send exactly the same helper data to the prover. More precisely, if the prover is an attacker and the helper sets sent to him in two requests differ in one point, then he has $K+1$ points that fit the polynomial, and he can compute the secret.
 In Subsection~\ref{subsec::DPAproof}, we propose a solution which does not need the mentioned conditions.
\end{remark}
\begin{remark}\label{remark::attacker_M}
The attacker only knows the helper data, $M$. He can use any arbitrary decoding function  $g(.)$ to estimate the secret, $\hat{S}$. His success probability is derived in the analysis of the soundness property in the following.
\end{remark}

Now, we show the $0$-completeness, $\frac{1}{|\mathcal{S}|}$-soundness and $0$-privacy properties of the proposed SS-FR2 protocol.\\
$\mathit{0}$\textit{-completeness}: The user $k$, having $(X_k,Y_k)$ and $M^{[k]}=\left\{(M_1,f(M_1)), \cdots , (M_K,f(M_K))\right\}$, has access to $K+1$ points of the polynomial, and thus he derives the polynomial with Lagrange interpolation and computes $a_0$ \cite{shamir1979share}. So, $\hat{S}^{[k]}=S^{[k]}=a_0$ for all $k\in [K]$ and thus:
\begin{align}
\mathbb{P}[\hat{\mathcal{H}}=1|\mathcal{H}=0]=\mathbb{P}[\hat{S}^{[k]}\neq S^{[k]}] =0. \label{eq::0complete_fr}
\end{align}
This proves the $0$-completeness property of the scheme.\\
$\mathit{\frac{1}{|\mathcal{S}|}}$\textit{-soundness}: Due to the information-theoretic secrecy of Shamir secret sharing, $M$ does not give any information about $S$ \cite{shamir1979share,fujiwara2016unbreakable}, i.e., $I(S;M)=0$. This means that if the prover is an attacker, by observing $M$ ($K$ points of the polynomial), the attacker obtains no information about the secret. 
Considering that $S$ is uniformly distributed, we have:
\begin{align}
\mathbb{P}[\hat{\mathcal{H}}=0|\mathcal{H}=1]=\mathbb{P}[\hat{S}= S]=\frac{1}{q^L}. \label{eq::sound_fr}
\end{align} 
This proves $\frac{1}{|\mathcal{S}|}$-soundness property and  
 means that the best possible attack strategy is to guess the secret.\\
$\mathit{0}$\textit{-privacy}: Every legitimate user can compute $S^{[k]}$ correctly. So, $\hat{S}^{[1]}=\hat{S}^{[k]}$ for all $k \in [K]$. Further, the authentication query is the same for all users.
 Thus, $(\hat{S}^{[1]},S^{[1]},M^{[1]},V,Q^{[1]})\sim(\hat{S}^{[k]},S^{[k]},M^{[k]},V,Q^{[k]})$ for all $k \in [K]$, and $0$-privacy is satisfied. \\
Now, we compute the normalized total key rate of the proposed SS-FR2 protocol as,
\begin{align}
R&=\!\frac{H(C_1,\!\cdots\!, C_K)}{K\text{length}(C)}\! =\!\frac{H((X_1,Y_1,Z_1),  \cdots , (X_K,Y_K,Z_K))}{Kl_cL} \nonumber \\
&\stackrel{(a)}=\frac{H(X_1,\cdots,X_K)+\sum_{i=1}^K H(Y_i)+\sum_{i=1}^K H(Z_i)}{Kl_cL} \nonumber \\
&\stackrel{(b)} =\frac{\log_q P(q^L-1,K)+K(l_c-1)L}{Kl_cL}, \label{R-fin}
\end{align}
where (a) follows from the fact that $Y_i$ and $Z_i$, for $i\in [K]$, are chosen independently and uniformly at random, and (b) is obtained by considering that $X_1, \cdots, X_K$ are chosen uniformly at random from all subsets of cardinality $K$ of the field $GF(q^L)$.
\qed

\begin{lemma} \label{lem::upp-R-fin}
The outer bound for $(R,R_c)$ pair ($R$ in $q$-ary unit) is equal to:
\begin{equation}
    \biggl\{ R_c\leq 1 ,\quad      R\leq 1\biggr\}.
\end{equation}
\end{lemma}

\begin{IEEEproof}
Noting \eqref{eq:rate}, we have:
\begin{align*}
R&=\frac{H(C_1,C_2,\cdots, C_K)}{K\text{length}(C)} \leq \frac{\sum_{i=1}^K H(C_i)}{KLl_c}=\frac{K L l_c}{KL l_c}=1.
\end{align*}
Indicating $R_c=\frac{1}{l_c}\leq 1$, the outer bound for $(R,R_c)$ is shown in Fig.~\ref{fig::thm1}.
\end{IEEEproof}

\subsection{Proof of Theorem \ref{asymptotic_thm} (single-server, asymptotic regime)} \label{subsec::asympsingle_proof}
We present the \emph{optimal} single-server PA protocol for the asymptotic regime.

\textbf{Converse}: Noting $C_k\!=\!Y_k^l$, it is obvious that $R \!\leq \!\max_{p(y)} H(Y)$.

\textbf{Achivability}:
The proof is based on the strong typicality \cite{DBLP:journals/tit/OrlitskyR01}.
We propose the achievable scheme for the single-server asymptotic regime (SS-AR) PA protocol, by defining the  functions at CA, the verifier, and the legitimate users.\\
1) \textit{Key distribution phase}:
First, we describe the codebook generation at CA. Considering auxiliary random variables $X \in \mathcal{X}$ and $U \in \mathcal{U}$, fix a joint distribution $P_{UXY}(u,x,y)=P_{X}(x) P_{Y|X}(y|x)$ $P_{U|X}(u|x)$, i.e., $U \to X \to Y$ forms a Markov chain, such that $0<I(U;Y) \leq I(X;Y)=\mu$.
Obtaining $P_U(u)=\sum_x P_{UX}(u,x)$ and fixing $\xi$ and $\xi^\prime$ such that $\xi > \xi^\prime >0$, form the set $\mathcal{J}$ by randomly and independently generating $2^{l\tilde{R}}$ sequences of $U^l$, each according to $\prod_{i=1}^l P_U(u_i)$.
Randomly partition the set $\mathcal{J}$ to $2^{lR^\prime}$ equal size subsets (bins).
So, each bin contains $2^{l(\tilde{R}-R^\prime)}$ sequences.
Each sequence $U^l \in \mathcal{J}$ can be shown as $U^l(m,s)$, where $m$ is the bin index and $s$ is the index of sequence in the bin.
The sequences $U^l(m,s)$ constitutes the codebook, which is revealed to every participant of the protocol (the verifier and both the legitimate users and attackers).\\
CA generates an i.i.d sequence $X^l$, from marginal distribution $P_X(x)$. Conditioned on $X^l$, randomly and conditionally independently, CA generates $K$  sequences $Y_1^l, Y_2^l, \cdots, Y_K^l$, each according to the conditional distribution $P_{Y|X}(y|x)$, i.e., $Y_k^l|X^l \sim \prod_{i=1}^l P_{Y|X}(y_i|x_i)$ for $k \in [K]$.
CA sets $C_k=Y_k^l$ and sends it to the user $k$.\\
2) \textit{Authentication phase}: considering user $k$ as the prover:

\textbf{S1:} The prover asks for authentication from the verifier.

\textbf{S2:} CA sets $V=\{X^l, Y_{1:K}^l\}$ and sends it to the verifier. 

\textbf{S3:} The verifier finds a sequence $U^l(m,s)\in \mathcal{J}$ that is jointly typical with $X^l$, i.e., $(X^l,U^l(m,s))\in \mathcal{T}^l_{\xi^\prime}(X,U)$. If no such sequence exists, the verifier chooses a sequence from $\mathcal{J}$ randomly.
If there is more than one sequence, the verifier chooses one of them randomly. Then, the verifier sets  $M^{[k]}=m$ and $S^{[k]}=s$. He sends $M^{[k]}$ to the prover.

\textbf{S4:} Having access to $Y_k^l$ and the bin index $M^{[k]}$, the prover looks for a sequence $\hat{U}^l$ in bin $M^{[k]}$ that is jointly typical with $Y_k^l$, i.e., $(\hat{U}^l(M^{[k]},\hat{s}),Y^l) \in \mathcal{T}_{\xi}^{(l)}(U,Y)$. One of the following cases occur:

$\bullet$ There is only one sequence in bin $M^{[k]}$ that is jointly typical with $Y^l$ $\longmapsto$ the prover takes this sequence as $\hat{U}^l$.

$\bullet$ There is more than one sequence in bin $M^{[k]}$ that are jointly typical with $Y^l$ $\longmapsto$ the prover chooses one of these sequences randomly and sets the selected sequence as  $\hat{U}^l$.

$\bullet$ There is no sequence in bin $M^{[k]}$ that is jointly typical with $Y^l$ $\longmapsto$ the prover chooses $\hat{U}^l$ at random from  bin $M^{[k]}$.\\
The index $\hat{s}$ of $\hat{U}^l$ in bin $M^{[k]}$ is the output of the decoding function, i.e., $ \hat{S}^{[k]}=\hat{s}$. The prover sends $\hat{S}^{[k]}$ to the verifier.

\textbf{S5:} The verifier checks if $\hat{S}^{[k]}=S^{[k]}$, to pass or reject the prover.

The above steps determine the variables of the  verification process $\mathcal{P}^{\text{single}}$ in the SS-AR protocol.

An outline of the analysis of the constraints follows. The detailed analysis is provided in Appendix~\ref{appthm2}.\\
For completeness, error occurs if $\hat{S}^{[k]} \neq S^{[k]}$.
By setting $\tilde{R}>I(U;X)$, there are more than $2^{l I(U;X)}$ sequences in $\mathcal{J}$. Using covering lemma \cite{el2011network}, with high probability the verifier finds a sequence $U^l$ that is jointly typical with $X^l$.
On the other hand, by the law of large number (LLN), the prover finds a sequence $\hat{U}^l$ that is jointly typical with $Y_k^l$. 
By setting $\tilde{R}-R^\prime<I(U;Y)$ and due to packing lemma, with high probability the prover finds a unique sequence. This proves that   $\lim_{l\to \infty} P_{e1}= 0$.\\
The soundness property is derived by showing that as $l\!\! \to\!\! \infty$, we have $\frac{1}{l} I(S;M)\!\!\to \!\! 0$  and using Lemma~\ref{lem1} in Appendix~\ref{applem}.\\
If $\{\hat{S}^{[1]} = \hat{S}^{[k]}\}$, privacy is guaranteed. Thus, $\mathbb{P}[\hat{S}^{[k]} \neq \hat{S}^{[k]}]$ is an upper bound on the probability of the event that privacy is violated. 
Utilizing a similar technique used in the proof of completeness property, we show that the probability of $\{\hat{S}^{[1]} \neq \hat{S}^{[k]}\} $ tends to zero. Since the number of legitimate provers is finite, $K$, the probability of $\{\hat{S}^{[1]} \neq \hat{S}^{[k]}\} $ for $\forall k \in [K]$ tends to zero.\\
Now, we analyze the normalized total key rate:
\begin{align*}
R&=\frac{H(C_1,C_2,\cdots, C_K)}{Kl}=\frac{H(Y_1^l,\cdots, Y_K^l)}{Kl} \nonumber \\
& \geq \frac{H(Y_1^l,\cdots, Y_K^l|X^l)}{Kl}  \stackrel{(a)}= \frac{1}{Kl} \sum_{i=1}^K  H(Y_i^l|X^l)  \stackrel{(b)}= H(Y|X),
\end{align*}
where (a) follows from the fact that $Y_1^l, Y_2^l, \cdots, Y_K^l$ are mutually independent conditioned on $X^l$ and (b) is derived by  $Y_k^l|X^l \sim \prod_{i=1}^l P_{Y|X}(y_i|x_i)$ for $k \in [K]$.

In the converse, we have shown that $R \leq \max_{p(y)} H(Y)$.
The difference of $H(Y)$ and $H(Y|X)$ is equal to $I(X;Y)=\mu$. If $\mu \to 0$ and by fixing $p(y)$ such that $H(Y)$ equals its maximum, then $\mathfrak{C}^{\text{asy}}= H(Y)$ is achievable. $\mu \to 0$ means negligible correlation between $X$ and $Y$, which results in vanishing correlation between the keys of the users.
\qed

\subsection{Proof of Theorem \ref{thm::Multi_server_fin} (multi-server, individual authentication, finite size regime) } \label{subsec::multi-indivi}
Here we consider the case of individual authentication, where 
the prover is connected to only one verifier and wants to be authenticated individually to that single verifier, who sends the helper data to the prover.
Other verifiers do not contribute in the process of authentication.
It must be emphasized that the prover has only one key, and uses this single key for authentication to any of the verifiers that he forms a feasible pair with.

Considering the feasible pair $(k,n^\ast)$, the decoding and encoding functions follow:\\
\textbf{scheme 1} of multi-server, finite size regime (MS-FR1) protocol ($R_c=1, R\leq \frac{1}{K}$):\\
1) \textit{Key distribution phase}: CA chooses $\tilde{C} \in \mathcal{W}=GF(q)^L$ uniformly at random and  sets $C_k=\tilde{C}$, $\forall k\in[K]$. The CA sends $C_k$ to user $k$.\\
2) \textit{Authentication phase}: For each feasible pair, the authentication phase is similar to  the authentication phase in scheme 1 (SS-FR1 protocol) in proof of Theorem~\ref{finitethm}. More precisely, CA sets $V_n=\tilde{C}$, the verifier $n^\ast$ sets $S_{n^\ast}^{[k]}=\tilde{C}$, and the prover (user $k$) sets $\hat{S}_{n^\ast}^{[k]}=C_k=\tilde{C}$.
The steps of the authentication phase specify the variables of $\mathcal{P}^{\text{Multi}}$. It is worth emphasizing that, $V_n$, $Q_n^{[k]}$,$S_n^{[k]}$ and $\hat{S}_{n}^{[k]}$ are null for $n\in [n]\backslash\{n^\ast\}$. $M_n^{[k]}$ is null $\forall n \in [N]$.

The analysis of completeness and soundness properties is similar to SS-FR1, equations~\eqref{eq::0complete_coomon_fr} and \eqref{eq::sound_common_fr}, respectively. Since every legitimate user sends exactly the same secret, $\tilde{C}$, privacy is preserved. The normalized total key rate can be computed according to \eqref{R-common-fin}. \\
\textbf{scheme 2} of multi-server, finite size regime (MS-FR2) protocol ($R_c=\frac{1}{2} (i\geq  2), R\leq 1-R_c+\frac{R_c}{KL} \log_q P(q^L-1,K)$): \\
1) \textit{Key distribution phase}: CA acts exactly the same as the key distribution phase in protocol SS-FR2, and sets the variables $X_{1:K}, Y_{1:K}, Z_{1:K}$. He sets $C_k=(X_k,Y_k,Z_k)$ and send $C_k$ to user $k$, $\forall k\in[K]$.\\
2) \textit{Authentication phase}: 

\textbf{S1:} The prover sends authentication query $Q^{[k]}$ to the verifier $n^\ast$. 

\textbf{S2:} CA chooses independently and uniformly at random $N$ variables $a_0^{(1)},a_0^{(2)},\cdots,a_0^{(N)}$ from the field $GF(q^L)$. He sets $V_n=\left\{a_0^{(n)}, C_k | k\in[K]\right\}$ and sends $V_n$ to the verifier~$n$, $\forall n\in [N]$. As stated in Remark \ref{remark::constant a0}, $a_0^{(1)},\cdots,a_0^{(N)}$ cannot be changed in different verification processes. Thus,  the CA can send $V_{1:N}$ to the verifiers once at the key distribution phase.

\textbf{S3:} The verifier $n^\ast$ sets $S^{[k]}_{n^\ast}=a_0^{(n^\ast)}$. 
The verifier $n^\ast$ computes unique polynomial $f_{n^\ast}(.)$ of degree $d_{n^\ast}=|\mathcal{N}_{n^\ast}|$ that passes through $\left\{(0,a_0^{(n^\ast)}), (X_k,Y_k)|k \in \mathcal{N}_{n^\ast}  \right\}$.
Then, he selects $d_{n^\ast}$ new points from $GF(q^L)$, that is the verifier chooses at random $d_{n^\ast}$ distinct elements $M_1^{(n^\ast)}, M_2^{(n^\ast)}, \cdots , M_{d_{n^\ast}}^{(n^\ast)}$ of $GF(q^L)\backslash\{0, X_k| \,  k\in \mathcal{N}_{n^\ast}\}$.
 The verifier computes the value of the polynomial at the new points and sets $M_{n^\ast}^{[k]}=\left\{(M_1^{(n^\ast)},f_{n^\ast}(M_1^{(n^\ast)})), \cdots,\right.$ $\left. (M_{d_{n^\ast}}^{(n^\ast)},f_{n^\ast}(M_{d_{n^\ast}}^{(n^\ast)}))\right\}$ and sends $M_{n^\ast}^{[k]}$ to user $k$.

\textbf{S4:} Knowing $M_{n^\ast}^{[k]}$ and $C_k$, the prover has access to $d_{n^\ast}\!+\!1$ distinct points of the polynomial $f_{n^\ast}(.)$ (of degree $d_{n^\ast}$), so he derives the polynomial, computes $\hat{S}_{n^\ast}^{[k]}=a_0^{(n^\ast)}$, and sends it to the verifier.

\textbf{S5:} The verifier checks if $\hat{S}_{n^\ast}^{[k]}=S_{n^\ast}^{[k]}$.

These steps determine the variables of the verification process $\mathcal{P}^{\text{Multi}}$.
Remarks~\ref{remark::constant a0} and \ref{remark::attacker_M} also apply here.\\
Considering \eqref{eq::0complete_fr} and \eqref{eq::sound_fr}, 0-completeness and $\frac{1}{|\mathcal{S}|}$-soundness are provided. Since every legitimate user can compute the secret correctly, \eqref{eq::privacy_MS} is satisfied with $P_P=0$ and 0-privacy is provided. $R=1-R_c+\frac{R_c}{KL} \log_q P(q^L-1,K)$ can be computed according to \eqref{R-fin}.
\qed

\begin{remark} \label{remark::MSPAFRintersection}
Consider the case that user $k_1$ and user $k_2$ form feasible pairs with both verifier $n_1$ and verifier $n_2$. Assume that user $k_1$ requests authentication to both verifier $n_1$ and verifier $n_2$. 
After the execution of the protocol MS-FR2, user $k_1$ computes polynomials $f_{n_1}$ and $f_{n_2}$. User $k_1$ knows that the keys of the users that are both in the $\mathcal{N}_{n_1}$ and $\mathcal{N}_{n_2}$ (e.g., user $k_2$) are parts of the intersection of the polynomials $f_{n_1}$ and $f_{n_2}$. This violates the security of the keys of users if the space of this intersection is small. This problem can be solved if the polynomial and the secret change in each execution of the authentication protocol. However, changing the secrets and as a result, the polynomials can lead to security violation (the detailed discussion is provided in Subsection~\ref{intro_extra}). 
The proposed protocol for the distributed version of the multi-server protocol in Subsection~\ref{subsec::DPAproof} solves this problem.
\end{remark}

\subsection{Proof of Theorem \ref{thm::DPA} (multi-server, distributed authentication, finite size regime)} \label{subsec::DPAproof}
Here, we consider the case of the multi-server scenario in which  all the verifiers contribute in the process of authentication, and so authentication is done in a distributed manner, i.e., all the verifiers contribute in sending helper data to the prover.
For this case, we propose an optimal interactive protocol, the distributed multi-server protocol in the finite size regime (DMS-FR), between the CA, the verifiers, and the prover that uses symmetric private information retrieval (SPIR) \cite{DBLP:journals/tit/SunJ19}, to deliver the prover the desired data, without realizing the identity of the prover.
 DMS-FR  is applicable in the condition that all the users are connecting to all the verifiers, i.e., $(k,n)$ forms a feasible pair $\forall k \in [K]$, $\forall n \in [N]$. 
For better understanding, first we present an example.
\begin{example}
Consider the case of two verifiers and three users ($N=2$, $K=3$), and
$\mathcal{S}=\mathcal{C}=GF(23)$, i.e, $q=23$ and $L=1$. User~$2$ wants to gain access to verifier $1$, and
the following protocol is executed: \\
1) \textit{Key distribution phase}: CA chooses uniformly at random from $GF(23)$ the variables $C_1=X_1=14$, $C_2=X_2= 19$ and $C_3=X_3=6$. These are the keys of the users that can be used in any verification process. \\
2) \textit{Authentication phase}:

$\bullet$ User $2$ sends an authentication query to all verifiers. In addition to authentication request to verifier $1$, user~$2$ sends queries to each verifier according to queries of SPIR (Appendix \ref{app::SPIR}) to retrieve his helper data.

$\bullet$ CA chooses randomly from $GF(23)$, $S=5$ as secret, $R=1$ as common randomness (for SPIR algorithm) and $\tilde{Y}=1$. He also chooses randomly $\tilde{X}=15$ from  $GF(23)\backslash\{0,X_1,X_2,X_3\}$.
CA sends $X_1$, $X_2$, $X_3$, $S$, $R$, and $(\tilde{X},\tilde{Y})$ to both verifiers.  

$\bullet$ Both verifiers compute the unique polynomial of degree one that passes through $(0,S)$ and $(\tilde{X},\tilde{Y})$, which is equal to $f(X)=12X+5$.
Then, both compute $Y_k=f(X_k)$ for $k\in \{1,2,3\}$. Thus, $Y_1=12$, $Y_2=3$ and $Y_3=8$. The verifiers store these variables. Then according to the requests that user~$2$ has sent to each verifier and the SPIR algorithm, using common variable $R$, the verifiers send answers to the user such that he can retrieve $Y_2=3$ privately without gaining any information about $Y_1$ and $Y_3$. Verifier 1, also, sends $(\tilde{X},\tilde{Y})=(15,1)$ to user $2$.

$\bullet$ User $2$ computes the unique polynomial that passes through $(X_2,Y_2)=(19,3)$ and $(\tilde{X},\tilde{Y})=(15,1)$, that is equal to $f(X)=12X+5$. Thus user $2$ can compute $S=f(0)=5$. He sends $S=5$ to verifier $1$.
\end{example}
\textbf{Converse}: According to Lemma~\ref{lem::upp-R-fin}, $R\le 1$ in $q$-ary unit and $R_c=\frac{1}{l_c}\leq 1$.

\textbf{Achievability}:
 We describe DMS-FR  for $l_c=1$, but any $l_c$ is achievable by enlarging the space of variables $l_c$ times.\\
Proof outline:
In one hand, the verifiers must not gain any information about the identity of the user by observing the requests for helper data (privacy constraint), and on the other hand, it is desired that the users cannot compute the keys of other users observing the helper data. Thus,
 the proposed DMS-FR protocol uses SPIR to privately deliver the desired data to the prover.
Here, the verifiers have the role of the servers with replicated data ($X_1, X_2, \cdots, X_K)$. In each verification process, the verifiers compute $Y_k=f(X_k),\, \forall k\in[K]$, where $f(.)$ is a degree one polynomial specified by the CA. So, $(Y_1, Y_2, \cdots, Y_K)$ are stored in all the verifiers. SPIR  is run between user~$k$ and the verifiers, such that the user can attain $Y_k$ privately, without gaining any information about $Y_{-k}$, where $Y_{-k}=(Y_1, Y_2, \cdots, Y_{k-1}, Y_{k+1}, \cdots ,Y_K)$. In addition to $Y_k$, another point on $f(.)$ is sent to the user as helper data. Using this point, $Y_k$, and $X_k$, known to the user as his personal key, the user can compute $f(.)$, and thus he can compute the secret. The detailed steps of the protocol are as follows:\\
1) \textit{Key distribution phase}: 
The CA chooses $X_1, \cdots, X_K$ from the field $\mathcal{C}=GF(q^L)$, independently and uniformly at random. It is worth noting that in this protocol, the condition that $X_k$'s must be distinct is not necessary.
Then, CA sends $X_k$ to the $k$-th legitimate user, $C_k=X_k$.\\
2) \textit{Authentication phase}:  in each verification process:

\textbf{S1:} User $k$ privately generates a random variable $\mathcal{F}$. This variable is not available to the verifiers and represents the randomness in the strategies followed by the user in the SPIR protocol \cite[Theorem 3]{DBLP:journals/tit/SunJ19} (described in Appendix~\ref{app::SPIR}).
For $n\in [N]$, user $k$ sends authentication query $Q_n^{[k]}$ to verifier $n$  according to SPIR,  to ask the verifiers to send $M_n^{[k]}$ for him in a manner to retrieve his specified data without revealing his identity\footnote{Considering the equivalence between $GF(q^L)$ and $\mathbb{F}_q^L$ SPIR can be executed.}. More precisely, user $k$ sends queries in a way that he can retrieve the desired data with index $k$, without revealing $k$ to any of the verifiers. His query to verifier $n^\ast$ also includes a request for authentication.

\textbf{S2:} The CA chooses $S$, randomly from the field $GF(q^L)$ as secret, and the point $(\tilde{X},\tilde{Y})$ where $\tilde{X} \in GF(q^L)\backslash\{0,X_k| \, k\in[K]\}$ and $\tilde{Y}\in GF(q^L)$ are chosen independently and uniformly at random.
 Also CA generates a common random variable $R\in \mathbb{F}_q$. CA  sends $\left\{X_{1:K}, S, R, (\tilde{X},\tilde{Y})\right\}$  to all the verifiers.

\textbf{S3:} Each verifier generates the unique polynomial $f(.)$ of degree one that passes through $\{(0,S), (\tilde{X},\tilde{Y})\}$ and computes $Y_k=f(X_k)$ for $k\in[K]$. 
 $Y_1\!,\! \cdots\!,\! Y_K$ are the replicated data stored in all the verifiers, $R$ is the common shared randomness between the verifiers and the prover wants to retrieve $Y_k$ privately. 
 According to $Q_n^{[k]}$  and $R$, using SPIR \cite[Theorem 3]{DBLP:journals/tit/SunJ19}, verifier $n$ generates answers $A_n^{[k]}$. The queries and the answers have the following properties:
\begin{align}
&H(A_n|Q_n, Y_1, \cdots, Y_K, R)=0  \label{eq::SPIR1}\\ 
&(\!Q_n^{[k_1]}\!,A_n^{[k_1]}\!,Y_{1:K}\!,R)\!\sim\!(\!Q_n^{[k]}\!,A_n^{[k]}\!,Y_{1:K}\!,R),\forall k_1,k\!\!\in\!\! [K]  \label{eq::SPIR2} \\
&I(Y_{-k};Q_{1:N}^{[k]},A_{1:N}^{[k]},\mathcal{F})=0, \forall k\in[K]. \label{eq::SPIR3}
\end{align}
The verifier $n^\ast$ sets $S_{n^\ast}^{[k]}=S$, and  sends $(\tilde{X},\tilde{Y})$ in addition to $A_{n^\ast}^{[k]}$ as helper data. More precisely, $M_n^{[k]}=A_n^{[k]}$ for $n\in[N]\backslash\{n^\ast\}$ and $M_{n^\ast}^{[k]}=\{A_{n^\ast}^{[k]},(\tilde{X},\tilde{Y})\}$.

\textbf{S4:} According to equation~\eqref{eq::SPIR-correctness}, user $k$ can compute $Y_k$ using $A_{1:n}^{[k]}$ correctly.
Thus, user $k$ knows $(\tilde{X},\tilde{Y})$ and $(X_k,Y_k)$. So, he can compute the polynomial $f(.)$ and thus he can find the secret.
The prover sends the estimated secret $\hat{S}_{n^\ast}^{[k]}$ to the verifier $n^\ast$.

\textbf{S5:} If $\hat{S}_{n^\ast}^{[k]}=S_{n^\ast}^{[k]}$ the verifier accepts the prover as a legitimate user.

These steps determine the variables of  $\mathcal{P}^{\text{Multi}}$.
Now we analyze completeness, soundness, and privacy conditions.\\
$\mathit{0}$\textit{-Completeness}: User $k$ has $X_k$ as his own personal key. In each verification process, he obtains $Y_k$ performing SPIR and $(\tilde{X},\tilde{Y})$ as a helper data. Thus, he has two points of a polynomial of degree one, and he can compute the polynomial and also the secret. So regardless of the verifier $\hat{S}_{n^\ast}^{[k]}=S_{n^\ast}^{[k]}$ for $k \in [K]$ and each legitimate user can compute the secret correctly. Thus:
\begin{align}
\mathbb{P}[\hat{\mathcal{H}}=1|\mathcal{H}=0]=\mathbb{P}[\hat{S}_{n^\ast}^{[k]}\neq S_{n^\ast}^{[k]}] =0.
\end{align}
This proves the $0$-completeness property of the scheme.\\
$\mathit{\frac{1}{|\mathcal{S}|}}$\textit{-Soundness}: In each verification process, the attacker can obtain $(\tilde{X},\tilde{Y})$ and $Y_k$, where $k$ is the index he masquerades with. But, since he has no information about $X_k$ (can be any element of the field $GF(q^L)$), he cannot 
gain any information about the secret, i.e., $I(S;M)=0$.
Considering that $S$ is uniformly distributed), we have:
\begin{align}
\mathbb{P}[\hat{\mathcal{H}}=0|\mathcal{H}=1]=\mathbb{P}[\hat{S}_{n^\ast}= S]=\frac{1}{q^L}. \label{eq::sound:DMSFR}
\end{align} 
This proves the $\frac{1}{|\mathcal{S}|}$-soundness property and means that the best possible attack strategy is guessing the secret. \\
$\mathit{0}$\textit{-Privacy}: Every legitimate user can compute the secret correctly and since SPIR protocol is performed (equation~\eqref{eq::SPIR2}) the verifiers cannot detect the index of the prover through queries they received, Thus $\forall k,k_1 \in [K]$ and $\forall n\in [N]$:
\begin{align*}
(V_n,Q_n^{[k_1]},S_n^{[k_1]},M_n^{[k_1]},\hat{S}_{n}^{[k_1]})\sim (V_n,Q_n^{[k]},S_n^{[k]},M_n^{[k]},\hat{S}_{n}^{[k]}),  
\end{align*}
and $0$-Privacy is satisfied.

Now we can compute the normalized total key rate of the proposed protocol in $q$-ary unit as:
\begin{align}
R&=\frac{H(C_1,C_2,\cdots, C_K)}{Kl} =\frac{H(X_1, X_2, \cdots , X_K)}{KL} \nonumber \\
&\stackrel{(a)}=\frac{H(X_1)+\cdots+ H(X_K)}{KL} =\frac{KH(X_1)}{KL}\stackrel{(b)}=1 ,\label{R-fin-multi}
\end{align}
where (a) and (b) are obtained by considering  $X_1, \cdots, X_K$ are chosen uniformly at random from the field $GF(q^L)$.
\qed 
\begin{remark}
In the DMS-FR protocol, any number (more than two) of the verifiers can participate (in the SPIR algorithm), but the larger the number of the verifiers the lower the total data rate between the verifiers and the prover  \cite{DBLP:journals/tit/SunJ19}. On the other hand, the smaller the number of the verifiers the lower the total data rate between the CA and the verifiers (because the CA sends $V_n$ to the smaller set of the verifiers).
\end{remark}

\subsection{Proof of Theorem \ref{thm::Multi_server_asymp} (multi-server, individual authentication, asymptotic regime)} \label{subsec::multi-asys}
Here, the multi-server asymptotic version of the PA protocol is considered, i.e., all the variables of the protocol are sequences of length $\Theta(l)$ for arbitrary large $l$. \\
\textbf{Converse}: For converse, it is obvious that $R \leq \max_{p(y)} H(Y)$.\\
\textbf{Achivability}:
The proof is mostly similar to the proof of Theorem~\ref{asymptotic_thm}.  The main difference is that the key of the user $k$ must be correlated to the data available at all the verifiers that he forms feasible pair with. 
The proposed achievable scheme in the multi-server in the asymptotic regime, (MS-AR) PA protocol, is as follows:\\
1) \textit{Key distribution phase}:
Considering auxiliary random variables $X_1, X_2, \cdots, X_N \in \mathcal{X}$ and $U_1, ..., U_N \in \mathcal{U}$, fix a joint distribution $P_{U_{1:N} X_{1:N} Y}(u_{1:N} ,x_{1:N} ,y)=$ $\prod_{n=1}^N \left( P_{X_n}(x_n)P_{U_n|X_n}(u_n|x_n)\right)  P_{Y|X_{1:N}}(y|x_{1:N})$, such that $0<I(U_n;Y) \leq I(X_{1:N};Y)=\mu , \, \forall n\in[N]$. Fixing $\xi$ and $\xi^\prime$ such that $\xi > \xi^\prime >0$ and for each $n \in [N]$,
obtaining marginal distributions $P_{U_n}(u_n)$, form the sets $\mathcal{J}_n$ by randomly and independently generating $2^{l\tilde{R}_n}$ sequences of $U_n^l$, each according to $\prod_{i=1}^l P_{U_n}(u_{n_i})$.
Randomly partition each set $\mathcal{J}_n$ to $2^{lR^\prime_n}$ equal size subsets (bins).
So, each bin contains $2^{l(\tilde{R}_n-R^\prime_n)}$ sequences.
Each sequence $U_n^l \in \mathcal{J}_n$ can be shown as $U_n^l(m_n,s_n)$, where $m_n$ is the bin index and $s_n$ is the index of sequence in the bin $m_n$.
The sequences $U_n^l(m_n,s_n)$ constitutes the codebooks, and are revealed to every participant of the protocol.
From each marginal distribution $P_{X_n}(x_n)$, CA generates i.i.d sequence $X_n^l$.
Denoting  $X_{\mathcal{N}^{[k]}}^l=\{X_n^l|n\in \mathcal{N}^{[k]}\}$. For each $k\in[K]$, generate the sequence $Y_k^l$ according to the conditional distribution 
$P_{Y|X_{\mathcal{N}^{[k]}}}$, i.e., $Y_k^l|X_{\mathcal{N}^{[k]}}^l \sim \prod_{i=1}^l P_{Y|X_{\mathcal{N}^{[k]}}}(y_i|x_{\mathcal{N}^{[k]}_i})$.
 CA sends $C_k=Y_k^n$ to the user~$k$ for $k \in [K]$.\\
2) \textit{Authentication phase}:
Verifier $n$ acts similar to SS-AR  using $U_n^l(m_n,s_n)$ as codebook.
 The prover proves that his key, $Y_k^l$, is correlated to the $n$-th verifier's sequence $X^l_n$ using codebook $U_n^l(m_n,s_n)$ and helper data $m_n$ that the verifier $n$ sends to him the same as SS-AR.

The analysis of probabilities of error is the same as SS-AR protocol, and for completeness the conditions $\tilde{R}_n>I(U_n;X_n)$ and  $\tilde{R}_n-R^\prime_n<I(U_n;Y)$ must be satisfied.\\
The normalized total key rate of the proposed scheme is:
\begin{align*}
&R=\frac{H(C_1,\cdots, C_K)}{Kl} =\frac{H(Y_1^l,\cdots, Y_K^l)}{Kl} \nonumber \\
& \geq \! \! \frac{H(Y_1^l,\dots, Y_K^l|X^l_{1:N})}{Kl}\! =\! \frac{\sum_{i=1}^K \!  H(Y_i^l|X^l_{1:N})}{Kl} \! =\! H(Y|X_{1:N}).
\end{align*}
In the converse we showed that $R \leq \max_{p(y)} H(Y)$.
The difference of $H(Y)$ and $H(Y\! |\! X_{1:N})$ is equal to $I(X_{1:N}\! ;\! Y)\!=\!\mu$. Thus, if $\mu \! \to\! 0$ and by fixing $p( y )$ such that $H(Y)$ reaches its maximum,  the optimal rate $\mathfrak{C}^{\text{asy}}\! =\! H(Y)$ is achievable.
\qed

\section{Comparison and concluding remarks} \label{Sec::Analyze&comp}
In this section, we first compare the proposed protocols and then we introduce some notable remarks.
\subsection{Comparison}
In this subsection, first, we compare the protocols in the finite size and asymptotic regimes.
Then, the differences and properties of individual authentication and distributed authentication are discussed.\\
\textbf{Finite size vs asymptotic}\\
In the finite size regime, the variables are elements of finite fields, but the variables in the asymptotic regime are sequences of length $\Theta(l)$, for arbitrary large $l$.
Although, all the protocols proposed for the finite size regime are practical, the SS-FR and MS-FR protocols are not optimal. Thus, the SS-AR and MS-AR protocols (in asymptotic regime) are proposed to show that optimal protocols are achievable.
\textbf{Individual vs distributed} \\
The main difference between these two authentication methods  is how the CA and the verifiers cooperate jointly to privately authenticate the prover.
In the individual case in each verification process, the prover is connected to only one verifier and that one verifier sends the helper data to the prover. Also, except the only time that the CA sends the keys of users (in the MS-FR1), the secret and the keys of the users (in the MS-FR2) and $X_n^l$ and the keys of the users (in the MS-AR) to the verifiers, there is no need for data exchange between the CA and the verifiers in any verification process. 
In the distributed case, the prover is connected to all (or some) of the verifiers and these verifiers send helper data to the prover. In the DMS-FR protocol, there is a permanent interaction between the CA and the verifiers in the process of authentication. In the following some other differences between MS-FR/MS-AR and DMS-FR are stated:

$\bullet$ DMS-FR is proposed for the case that all the users are making feasible pairs with all the verifiers, however MS-FR and MS-AR protocols are feasible for any connection structure.

$\bullet$ As stated in Remark~\ref{remark::MSPAFRintersection}, in the MS-FR2 protocol, the users can gain information about other users' keys (the space of the estimated keys may be small). Thus, security of the keys is not provided. However, in the DMS-FR, because of the changes of the polynomial and the secret in each verification process, the users cannot gain any information about other users' keys. This happens because if a user wants to gain information about other users' keys, he acts exactly the same as an attacker, which his functionality is analyzed in soundness property in Theorem~\ref{thm::DPA}.

$\bullet$ The size of the keys are smaller in the DMS-FR compared with the MS-FR2 protocol for a fixed $P_{e2}$-soundness constraint. For example for  $P_{e2}=\frac{1}{q^L}$, in DMS-FR $C_k\in GF(q^L)$, but in MS-FR2, $C_k \in GF(q^L)^2$.

$\bullet$ The complexity of computing secret in the DMS-FR protocol is far less than the MS-FR2 protocol. Since in the DMS-FR protocol, the participants compute a polynomial of degree one, but in the MS-FR2 protocol, the polynomial of verifier $n$ is of degree equal to $|\mathcal{N}_n|$.
\begin{remark}
Some extra services are available in  DMS-FR; however, none of the MS-FR1, MS-FR2, and MS-AR can provide these services. The detailed discussion is stated in Subsection~\ref{intro_extra}.
\end{remark}

\subsection{Extra Services} \label{intro_extra}
Some extra services can be considered in PA protocol, which we describe in the following. While these services are not central to the protocol, they can strengthen the performance of the protocol. We discuss which of the proposed protocols in Section \ref{sec::proofs} provide each extra service.\\
\textbf{Auditing:} This service enables the CA to recognize the prover.
There are systems and services that the CA is required to recognize the prover, e.g., to identify the malicious behavior. More precisely, auditing means while the verifier (verifiers) cannot distinguish the index of the prover (i.e., legitimate user), the CA can distinguish the index of the legitimate user by observation of the verification process. Noting that the CA observes the entire process and $C, Q_{1:N}^{[\theta]}, S_{1:N}^{[\theta]}, M_{1:N}^{[\theta]}, \hat{S}_{1:N}^{[\theta]},V$ variables, this means,
\begin{align}
H(\theta|C, Q_{1:N}^{[\theta]}, S_{1:N}^{[\theta]}, M_{1:N}^{[\theta]}, \hat{S}_{1:N}^{[\theta]},V_{1:N} )=0, \label{eq::audit}
\end{align} 
where $\theta$ is uniformly and randomly chosen from $[K]$.

$\bullet$ \textit{SS-FR1, SS-FR2, SS-AR, MS-FR1, MS-FR2 and MS-AR}: In these protocols, each legitimate prover sends the same secret key in any verification process. Moreover, the authentication query is just a request for authentication and thus, there is no difference between the users. So, neither the verifier nor the CA can detect the ID of the legitimate prover. Thus auditing is not provided.

$\bullet$ \textit{DMS-FR}: Since CA has access  to all of the verifiers, he has access to $Q_{1:N}^{[k]}$, considering the prover is user~$k$. 
More precisely, in SPIR, if one has access to all the queries sent from the prover he can recognize the index of the data the prover asks for. Thus, the CA can recognize the index $"k"$ and  \eqref{eq::audit} is satisfied.\\
\textbf{Addition and deletion of users:} This service means that the set of users, $[K]$, and its partitioning in the multi-server scenario can be changed in each verification process. In general, this service enables the CA to update the list of users (add or delete a user) and its partitioning in each verification process without information leakage (to the attackers or other users) of secret variables of the protocol.

$\bullet$  \textit{SS-FR1 and MS-FR1}: It is possible to add a new user, by sending the common key, $\tilde{C}$, to him. Deleting a user is only possible by re-running the key distribution phase and generating a new $\tilde{C}$, such that the deleted user does not have access to it.

$\bullet$ \textit{SS-FR2 and MS-FR2}:
This property cannot be provided. Adding and deleting a user change the unique polynomial that passes through the keys of the legitimate users. The intersection points of the polynomial before deleting or adding a user and the new polynomial are the keys $(X_k,Y_k)$ of the other users. Thus, this can reveal the keys. Of course, only legitimate users can compute the polynomials, and as a result the keys of other users.

$\bullet$ \textit{SS-AR and MS-AR}: It is possible to add a new legitimate user, user $K+1$, by generating $Y_{K+1}^n$ using appropriate conditional distribution  the same as the process of generating other keys (key distribution phase in the proof of Theorems~\ref{asymptotic_thm} and \ref{thm::Multi_server_asymp}). 
But for the deletion of a user, the entire key distribution phase must  be re-run to generate new $X^n$ and new keys for other users. As a result, the key of the deleted user is not jointly typical with new $X^n$ (with high probability), and thus the deleted user cannot compute the new secret.
 
 $\bullet$ \textit{DMS-FR}: Both addition and deletion of users can be provided in this protocol. Since the  degree one polynomial function, $f(.)$, in each verification process does not depend on the keys of users, addition and deletion of the keys do not leak any information.
In other words, if CA wants to add a user to the protocol, he should select a key randomly, update the list of  the keys, and announce this update to all the verifiers. On the other hand, if the CA wants to delete a user, he declares to the verifiers to delete that user's key. \\
\textbf{Expiring secret:} This condition implies that the secret must be changed in each verification process. The reason behind this is that the legitimate user might be intended to reveal the secret since their own personal keys are immune. In addition, in the verification process or secret storing, the secret might be exposed.
 Thus, if in each verification process, the secret changes, the security of the protocol is enhanced.

$\bullet$  \textit{SS-FR1 and MS-FR1}: Since the secret is equal to $\tilde{C}$, key of all users, it is not possible to change the secret in each verification process.

$\bullet$  \textit{SS-FR2 and MS-FR2}: Changing the secret in each verification process means changing the unique polynomial, similar to the condition in addition and deletion service, and by detecting the intersection points of these polynomials the users can get $(X_k,Y_k)$ of other users. So this property is not provided.

$\bullet$ \textit{SS-AR and MS-AR}:  In this protocol changing the secret in each  verification process means changing $X^n$ and as a result changing the keys of the users. So, the expiring secret feature is not provided.

$\bullet$ \textit{DMS-FR}: As stated in the protocol in each verification process the CA chooses a secret at random. So, the secret changes in every verification process and this feature is available.
\subsection{Additional Attacks} \label{intro_attack}
We consider the following attacks in addition to two attacks that are considered in the system framework, 1) the (semi-honest) verifier wants to violate privacy, 2) an attacker wants to be authenticated illegally.
 We analyze the resistance of our proposed protocols against these attacks.\\
\textbf{Colluding legitimate users:} Legitimate users collude to gain information about other users' keys.

$\bullet$ \textit{SS-FR1 and MS-FR1}: Since the keys of all users are the same, this attack is not valid.

$\bullet$ \textit{SS-FR2 and MS-FR2}: Colluding legitimate users cannot get any information beyond the unique polynomial. So they cannot get additional information about other users' keys by colluding, and the protocol is resistant against this attack.

$\bullet$ \textit{SS-AR and MS-AR}:  Since the keys of the legitimate users have some common information, colluding users can gain some information about other users' keys. 

$\bullet$ \textit{DMS-FR}:  Each authentication request means a new verification process and as a result, a new secret and a new polynomial. Thus colluding legitimate users have access to secret and polynomials that are independent from each other, so they cannot get any additional information by colluding. If a legitimate user wants to attack the protocol to violate secrecy, his performance is like an attacker, that his success probability is computed in \eqref{eq::sound:DMSFR}.\\
\textbf{Malicious Verifier:} We have considered semi-honest verifiers, who follow the protocol. But, in a stronger attack scenario, it is possible that the verifiers do not follow the protocol and send helper data in a way to violate the privacy of the prover or disrupt the process of authentication. In this situation, the  verifier may send incorrect helper data to distinguish the identity of the legitimate prover or disrupt the authentication process.

$\bullet$ \textit{SS-FR1 and MS-FR1}: No helper data is sent in these schemes, thus, this attack is not valid and cannot be occurred.

$\bullet$ \textit{SS-FR2, and MS-FR2}: There are two possibilities for the malicious verifier action:
\begin{enumerate}
\item The malicious verifier can include the key of one or some of the users in the helper data. More precisely, $C_{k_1}, C_{k_2}, \cdots,C_{K_j} \subset C$ can be part of helper data. In this case, if the prover is one of $K_1, K_2, \cdots, K_j$, he has $K$ points of the polynomial, so he cannot compute the polynomial and cannot be authenticated as a legitimate user. However, from the viewpoint of the verifier, there is no difference between this user and an attacker. So, the verifier cannot violate the privacy of the user. Of course, the user realized that the verifier does not follow the protocol and can report. On the other hand, if the prover is not any of $K_1, K_2, \cdots, K_j$, authentication continues and the verifier realizes that the prover is one of the $[K]\backslash \{K_1, K_2, \cdots, K_j\}$.
\item The malicious verifier can send completely wrong helper data. In this case, the verifier can check that with  which key  the estimated secret is computed and detect the user.
\end{enumerate}
These protocols cannot overcome this attack, unless CA himself  produces helper data, sends the helper data to the verifier and forces the verifier to send this helper data.

$\bullet$ \textit{SS-AR and MS-AR}: This means that the verifier sends the wrong bin index. These protocols cannot overcome this attack, unless CA himself produces helper data, sends the helper data to the verifier and forces the verifier to send the correct helper data.

$\bullet$ \textit{DMS-FR}: If the malicious verifier is any verifier except $n^\ast$, since the estimated secret is not sent to him, he cannot detect the index of the user and violate privacy. However, the prover cannot estimate the secret correctly. If the malicious verifier is $n^\ast$, he can send $\bar{X}$ and $\bar{Y}$ in such a way to distinguish the index of user and privacy is violated. The protocol cannot overcome any of these situations.\\
\textbf{Colluding verifiers:} Verifiers collude to detect the prover's identity and violate privacy. It is worth mentioning that this attack is possible in the multi-server scenario.

$\bullet$ \textit{MS-FR1, MS-FR2, and MS-AR}: In these protocols, in each verification process only one verifier is participating. Thus, this attack is not feasible.

$\bullet$ \textit{DMS-FR}: Colluding verifier can get some information about the index of the data requested by the prover and thus, they may gain some information about the prover's identity. To overcome this attack a $T$-collude SPIR algorithm according to \cite{DBLP:conf/allerton/WangS17} can be performed, which is resistant against colluding of $T$ verifiers.
\subsection{Distributed Authentication for any Structure} \label{conclude}
 The proposed DMS-FR PA protocol is feasible for the cases that all the verifiers are connected to all the provers. Solving the distributed authentication problem in the multi-server PA protocol scenario for any desired connection structure between the provers and the verifiers is of interest.
Of course, in the case that for each prover, there exist at least two verifiers such that the prover forms feasible pair with them, $ \mathcal{N}^{[k]}$, and $\forall k \in [K]$, we have $A^{[k]}=\{\bigcap_{n\in \mathcal{N}^{[k]}} \mathcal{N}_n\} \backslash \{k\} \neq \emptyset$, similar protocol to DMS-FR can be used as a solution. In the solution, when user $k$ requests authentication, only the verifiers in $\mathcal{N}^{[k]}$ collaborate in the process of authentication (especially the SPIR algorithm). In this case, considering user $k$, privacy is satisfied between $k$ and provers in $A^{[k]}$.
\subsection{Key-sharing}
In the finite size regime, by repeating the verification process $t$ times to improve security (soundness), then the achievable region can be defined in a real-valued alphabet. 
Then asymptotically using key-sharing, the lower convex of the achievable points can be achieved.
Inspired by time-sharing, we propose key-sharing scheme, where if any two points $(R_1,{R_c}_1)$ and $(R_2,{R_c}_2)$ are achievable, then the line connecting them is achievable, too.
\appendices
\section{} \label{applem}
\begin{lemma} \label{lem1}
If $I(U;J)\leq \alpha$, where $U \in \mathcal{U}$ and $J \in \mathcal{J}$, then $\mathbb{P}[U=g(J)] \leq \frac{1+\alpha+\log |\mathcal{U}|-H(U)}{\log |\mathcal{U}|} $, with $g: \mathcal{J}\to \mathcal{U}$ be any arbitrary function.
\end{lemma}
\begin{IEEEproof}
Let $\hat{U}=g(J)$. Using Fano's inequality, we have
\begin{align}
H(U|\hat{U}) \leq 1+\mathbb{P}[U\neq \hat{U}] \log |\mathcal{U}|. \label{fano}
\end{align} 
On the other hand, since $U\to J \to \hat{U}$ forms a Markov chain and $I(U;J)\leq \alpha$, we obtain $I(U;\hat{U})\leq \alpha$, and thus:
\begin{align}
H(U|\hat{U}) \geq H(U)-\alpha. \label{eq::markov}
\end{align}
Combining \eqref{fano} and \eqref{eq::markov}, we have:
\begin{align*}
H(U) - \alpha\leq 1+\mathbb{P}[U\neq \hat{U}] \log |\mathcal{U}|,
\end{align*}
and noting that  $1-\mathbb{P}[U\neq \hat{U}]=\mathbb{P}[U= \hat{U}]$, we obtain:
\begin{align*}
 \mathbb{P}[U= \hat{U}] \leq \frac{1+\alpha+\log |\mathcal{U}|-H(U)}{\log |\mathcal{U}|}.
\end{align*}
\end{IEEEproof}

\section{Probability of error analysis of Theorem \ref{asymptotic_thm}} \label{appthm2}
To complete the proof, here we have to show that in the proposed SS-AR protocol $P_{e1}$, $P_{e2}$, and $P_P$ tends to zero as $l$ goes to infinity.\\
\textit{Completeness}: The verifier rejects ($\hat{\mathcal{H}}=1$) the legitimate prover (user $k$, $\mathcal{H}=1$) if $\hat{S}^{[k]} \neq S^{[k]} $, and this may happen if either of the following events occurs:
\begin{enumerate}
\item $E_1$: The verifier does not find a sequence $U^l$ that is jointly typical with $X^l$.
\item $E_2$: User $k$ does not find a sequence $\hat{U}^l$ that is jointly typical with $Y_k^l$.
\item $E_3$: User $k$ finds more than one sequence $\hat{U}^l$  in bin $M^{[k]}$ that is jointly typical with $Y_k^l$.
\end{enumerate} 
Using the union bound, we obtain:
\begin{align*} \label{comp}
\mathbb{P}[\hat{\mathcal{H}}=1|\mathcal{H}=0]=\mathbb{P}[\hat{S}^{[k]} \neq S^{[k]} ] &\leq \mathbb{P}[E_1]+\mathbb{P}[E_2 \cap E_1^c] \nonumber \\
&+\mathbb{P}[E_3\cap E_1^c]. 
\end{align*}
By covering lemma \cite{el2011network},
if $\tilde{R}\geq I(X;U)+\varphi(\xi^\prime)$ (there are more than $2^{l I(U;X)}$ sequences in $\mathcal{J}$), $\mathbb{P}[E_1] \to 0$ when $l \to \infty$.
 Also, using LLN and the fact that $\xi>\xi^\prime$, $\mathbb{P}[E_2 \cap E_1^c] \to 0$ as $l\to \infty$. 
 And if $\tilde{R}-R^\prime \leq I(U;Y)-\varphi(\xi)$ (there exist less than $2^{l I(U;Y)}$ sequences in bin $M^{[k]}$), according to packing lemma \cite{el2011network}, $\mathbb{P}[E_3 \cap E_1^c] \to 0$ for large enough $l$. 
$\varphi(\xi)$ and $\varphi(\xi^\prime)$ are functions of $\xi$ and $\xi^\prime$, respectively, that goes to $0$ as $\xi,\xi^\prime \to 0$.
So, by setting $\tilde{R}=I(X;U)+\varphi(\xi^\prime)$ and $R^\prime=I(X;U)-I(U;Y)+\varphi(\xi)+\varphi(\xi^\prime)$, we obtain:
\begin{align}
\lim_{l\to \infty} \mathbb{P}[\hat{\mathcal{H}}=1|\mathcal{H}=0]=0.
\end{align}
This proves the completeness property of the proposed scheme.\\
\textit{Soundness}: From \cite[(31)]{lai2008privacy}, we have $I(S;M)\leq l(\varphi(\xi^\prime) + \zeta)$, where $\zeta$ is a function of $\xi^\prime$ and goes to $0$ as $l \to \infty$. This means that the data the verifier sends to the prover contains negligible information about the secret on average. Substituting $\alpha= l(\varphi(\xi^\prime) + \zeta)$ in Lemma~\ref{lem1}, we conclude that $\mathbb{P}[S= \hat{S}] \leq \frac{1+ l(\varphi(\xi^\prime) + \zeta+\log |\mathcal{S}|-H(S))}{\log |\mathcal{S}|}$. Since $S$ ranges from $1$ to $2^{l(I(U;Y)-\varphi(\xi))}$, we have $\log |\mathcal{S}|=l(I(U;Y)-\varphi(\xi))$ and
using \cite[(30)]{lai2008privacy}, $H(S) \geq l\left(I(U;Y)-\varphi(\xi)-\varphi(\xi^\prime)-\zeta\right)$.
 Thus:
\begin{align}
\mathbb{P}[\hat{\mathcal{H}}=0|\mathcal{H}=1]=\mathbb{P}[S= \hat{S}] \leq \frac{1+ l(2 \zeta + 2 \varphi(\xi^\prime))}{l(I(U;Y)-\varphi(\xi))}.
\end{align}
Since  $\xi, \xi^\prime \to 0$, we have $\lim_{l \to \infty} \frac{(2 \zeta + 2 \varphi(\xi^\prime))}{(I(U;Y)-\varphi(\xi))} \to 0$, and the soundness property is proved.\\
\textit{Privacy}: Now we analyze $P_p$-privacy property of the suggested protocol. First, we define the event $E$ as:
\begin{align*}
E=&\left[ (V,Q^{[1]},S^{[1]},M^{[1]},\hat{S}^{[1]})\sim \right.\\
&\left.(V,Q^{[k]},S^{[k]},M^{[k]},\hat{S}^{[k]}),\forall k \in [K] \right].
\end{align*}
If event $E$ occurs with probability $1$, we have perfect privacy. Now we consider the case that $E$ does not occur.
\begin{align}
\mathbb{P}[E^c]&=\mathbb{P}\left[\exists k \in [K]:\, (V,Q^{[1]},S^{[1]},M^{[1]},\hat{S}^{[1]})\nsim \right.\nonumber  \\ &\left.(V,Q^{[k]},S^{[k]},M^{[k]},\hat{S}^{[k]}) \right]\nonumber  \\
 &\stackrel{(a)}\leq \mathbb{P}[\exists k \in [K]:\, \hat{S}_1 \neq \hat{S}_k] \stackrel{(b)} \leq \sum_{k=2}^K \mathbb{P}[\hat{S}_1 \neq \hat{S}_k ],
\end{align}
where (a) follows from the fact that if $(V,Q^{[1]},S^{[1]},M^{[1]},\hat{S}^{[1]})\nsim (V,Q^{[k]},S^{[k]},M^{[k]},\hat{S}^{[k]})$, then $\hat{S}^{[1]} \neq \hat{S}^{[k]} $ (while there may exist situations that $\hat{S}^{[1]} \neq \hat{S}^{[k]} $ but $(V,Q^{[1]},S^{[1]},M^{[1]},\hat{S}^{[1]})\sim (V,Q^{[k]},S^{[k]},M^{[k]},\hat{S}^{[k]})$),
and (b) is obtained using union bound.
Now, for a fixed $k$ we compute $ \mathbb{P}[\hat{S}^{[1]} \neq \hat{S}^{[k]} ]$.
$\{\hat{S}^{[1]} \neq \hat{S}^{[k]}\}$ may happen in the following cases:
\begin{itemize}
\item $P_1$: The verifier does not find a sequence $U^l$ that is jointly typical with $X^l$.
\item $P_2$: Both user $1$ and user $k$ do not find a sequence $\hat{U}^l$ that is jointly typical with $Y_1^l$ and $Y_k^l$, respectively.
\item $P_3$: User $1$ finds a unique $\hat{U}^l$ that is jointly typical with $Y_1^l$, but user $k$ does not find a sequence $\hat{U}^l$ that is jointly typical with $Y_k^l$.
\item $P_4$: User $k$ finds a unique $\hat{U}^l$ that is jointly typical with $Y_k^l$, but user $1$ cannot find a sequence $\hat{U}^l$ that is jointly typical with $Y_1^l$.
\item $P_5$: User $1$ finds a unique $\hat{U}^l$ that is jointly typical with $Y_1^l$, but user $k$ finds more than one sequence $\hat{U}^l$ that is jointly typical with $Y_k^l$.
\item $P_6$: User $1$ does not find a sequence $\hat{U}^l$ that is jointly typical with $Y_1^l$, but user $k$ finds more than one sequence $\hat{U}^l$ that is jointly typical with $Y_k^l$.
\item $P_7$: User $k$ finds a unique $\hat{U}^l$ that is jointly typical with $Y_k^l$, but user $1$ finds more than one sequence $\hat{U}^l$ that is jointly typical with $Y_1^l$.
\item $P_8$: User $k$ does not find a sequence $\hat{U}^l$ that is jointly typical with $Y_k^l$, but user $1$ finds more than one sequence $\hat{U}^l$ that is jointly typical with $Y_1^l$.
\item $P_9$: Both user $1$ and user $k$ find more than one sequence $\hat{U}^l$ that is jointly typical with $Y_1^l$ and $Y_k^l$, respectively.
\end{itemize}
Using union bound, we have:
\begin{align}
\mathbb{P}[\hat{S}^{[1]} \neq \hat{S}^{[k]} ] \leq \sum_{i=1}^9 \mathbb{P}[P_i].
\end{align}
It should be considered that there are $2^ {l(I(U;X)+\varphi(\xi^\prime))}$ of $U^l$ sequences in $\mathcal{J}$ and there are $2^{l(I(U;Y)-\varphi(\xi))}$ of $U^l$ sequences in each bin.
As a result,
by covering lemma, $ \mathbb{P}[P_1]\to 0$. By LLN, $ \mathbb{P}[P_2]$, $ \mathbb{P}[P_3]$, and $ \mathbb{P}[P_4]$ go to zero when $l$ is sufficiently large. By packing lemma and LLN, $ \mathbb{P}[P_i]$ for $i\in\{5,6,7,8,9\}$ tends to zero as $l \to \infty$. So, we conclude:
\begin{align}
\lim_{l\to \infty}\mathbb{P}[\hat{S}^{[1]} \neq \hat{S}^{[k]} ] =0,
\end{align}
and since $K$ is fixed:
\begin{align*}
\lim_{n\to \infty}\mathbb{P}[E^c]=0,
\end{align*}
and
\begin{align}
\lim_{l\to \infty}\mathbb{P}&\left[ (V,Q^{[1]},S^{[1]},M^{[1]},\hat{S}^{[1]})\sim \right. \nonumber \\
&\left.(V,Q^{[k]},S^{[k]},M^{[k]},\hat{S}^{[k]}) \quad \forall k \in [K] \right] =1.
\end{align}
This completes privacy property.
\qed

\section{Symmetric Private Information Retrieval } \label{app::SPIR}
Here, we describe the SPIR protocol presented in \cite{DBLP:journals/tit/SunJ19} and is used in our proposed DMS-FR protocol.

\begin{figure}
\centering
\scalebox{0.6}{%

\begin{tikzpicture}

\draw[fill=blue!10] (2,5) rectangle (4,10) node[pos=.5]{$\begin{aligned} Y_1 \\ Y_2 \\ \cdots \\ Y_K \\ R \end{aligned}$};

\draw[fill=orange!30] (7,5) rectangle (9,10) node[pos=.5]{$\begin{aligned} Y_1 \\ Y_2 \\ \cdots \\ Y_K \\ R \end{aligned}$};
\draw[fill=green!50] (4.5,3) rectangle (6.5,4) node[pos=.5]{User};

\node at (3,10.5) {DB1};
\node at (8,10.5) {DB2};

\node at (1.5,7) {\Huge $\theta$};
\node at (9.5,7) {\Huge $\theta$};

\node [red] at (1,7) {\Huge  \ding{56}};
\node [red] at (10,7) {\Huge \ding{56}};

\path (6.3,4.01) node (U2) {};
\path (4.7,4.01) node (U1) {};
\path (3,4.99) node (B1) {};
\path (8,4.99) node (B2) {};
\draw [->,thick] (B1) -- node [below, midway] {Answer} (U1);
\draw [->,thick] (B2) -- node [below, midway] {Answer} (U2);

\path (5.8,4.01) node (U22) {};
\path (5.2,4.01) node (U11) {};
\path (3.5,4.99) node (B11) {};
\path (7.5,4.99) node (B22) {};
\draw [->,thick] (U11) -- node [above, pos=0.4] {Query} (B11);
\draw [->,thick] (U22) -- node [above, pos=0.1] {Query} (B22);

\node at (5.5,2.5) {\Large $Y_\theta$};
\node [green] at (6.1,2.5) {\Large \ding{52}};
\node at (5.5,2) {\Large $Y_1, \cdots, Y_{\theta-1}, Y_{\theta+1}, \cdots, Y_K$};
\node [red] at (8.5,2) {\Large \ding{56}};

\end{tikzpicture}
}
\caption{The SPIR problem with two databases and $K$ messages.}
\label{fig::SPIR}
\end{figure}  

Consider $K$ independent messages, $Y_1, Y_2, \cdots, Y_K$, and $N$ servers that each stores all the messages.
 The user $\theta$ wishes to retrieve $Y_\theta$ privately, and the servers do not want the user to get any information beyond the desired message, $Y_\theta$ (database privacy).
The user generates a random variable $\mathcal{F}$  privately. This random variable  represents the randomness in the strategies followed by the user, and its realization  is not available to the servers.
The servers share a common random variable $R$ that is unknown to the user, to achieve database privacy. An example of the SPIR problem is illustrated in Fig.~\ref{fig::SPIR}. 

In order to retrieve message $Y_k, \, k\in [K]$ privately, the user privately generates $N$ queries $Q_1^{[k]}, \cdots, Q_N^{[k]}$,
\begin{equation}
H(Q_1^{[k]}, \cdots, Q_N^{[k]}|\mathcal{F})=0.
\end{equation}
The user sends $Q_n^{[k]}$ to the $n$-th server, $n \in [N]$. Upon receiving $Q_n^{[k]}$, the $n$-th server generates an answer sequence $A_n^{[k]}$, which is a function of $Q_n^{[k]}$, all messages $Y_1, \cdots , Y_K$, and the common randomness $R$,
\begin{equation}
H(A_n^{[k]}|Q_n^{[k]},Y_1, \cdots , Y_K, R)=0.
\end{equation}
The $n$-th server returns $A_n^{[k]}$ to the user.
The user decodes $Y_k$ using the information available to him, $A_{1:N}^{[k]},Q_{1:N}^{[k]},\mathcal{F}$, according to a decoding rule specified by the SPIR scheme, that is:
\begin{equation}
\frac{1}{L} H(Y_k|A_{1:N}^{[k]},Q_{1:N}^{[k]},\mathcal{F})=o(L), \label{eq::SPIR-correctness}
\end{equation}
where $H(Y_k)=Ll_k, \, k\in[K]$.

To protect the users' privacy, K strategies must be indistinguishable from the perspective of any individual server:
\begin{align}
&(A_{1:N}^{[k]},Q_{1:N}^{[k]},Y_{1:K},R)\sim \nonumber \\
&(A_{1:N}^{[k^\prime]},Q_{1:N}^{[k^\prime]},Y_{1:K},R) \quad \forall k, k^\prime \in [K], \forall n \in [N].
\end{align}

To protect the privacy of database, the user must not get any information beyond his desired message, thus:
\begin{equation}
I(Y_{-k};A_{1:N}^{[k]},Q_{1:N}^{[k]},\mathcal{F})=0, \quad \forall k \in [K],
\end{equation}
where $Y_{-k}=Y_1, \cdots, Y_{k-1}, Y_{k+1}, \cdots, Y_K$.

For detailed description of the SPIR problem and its scheme, please refer to \cite{DBLP:journals/tit/SunJ19}.

\bibliographystyle{./IEEEtran}
\bibliography{./IEEEabrv,./paper}
\end{document}